\documentclass[preprint,letterpaper,prc,aps,showpacs,showkeys]{revtex4}
\usepackage{graphicx}
\usepackage{epsfig}

\newcommand{\beq}{\begin{equation}}
\newcommand{\eeq}{\end{equation}}
\newcommand{\beqa}{\begin{eqnarray}}
\newcommand{\eeqa}{\end{eqnarray}}

\newcommand{\Sigs}{\Sigma_{\mathrm s} }
\newcommand{\Sigv}{\Sigma_{\mathrm v} }
\newcommand{\Sigo}{\Sigma_{\mathrm 0} }

\newcommand{\kf}{k_{\mathrm F} }

\newcommand{\bfgamma}{\mbox{\boldmath$\gamma$\unboldmath}}
\newcommand{\bftau}{\mbox{\boldmath$\tau$\unboldmath}}

\newcommand{\veck}{\textbf{k}}

\newcommand{\vecq}{\textbf{q}}

\newcommand{\pabs}{|{\bf p}|}
\newcommand{\qabs}{|{\bf q}|}
\newcommand{\kabs}{|{\bf k}|}
\makeatletter
\newcommand{\fmslash}[2][0mu]{%
  \mathchoice
    {\fmsl@sh\displaystyle{#1}{#2}}%
    {\fmsl@sh\textstyle{#1}{#2}}%
    {\fmsl@sh\scriptstyle{#1}{#2}}%
    {\fmsl@sh\scriptscriptstyle{#1}{#2}}}
\newcommand{\fmsl@sh}[3]{%
  \m@th\ooalign{$\hfil#1\mkern#2/\hfil$\crcr$#1#3$}}
\makeatother

\begin{document}
\preprint{}
\title{The relativistic self-energy in nuclear dynamics} 
\author{O. Plohl}
\author{C. Fuchs}
\affiliation{Institut
f$\ddot{\textrm{u}}$r Theoretische Physik, Universit$\ddot{\textrm{a}}$t
T$\ddot{\textrm{u}}$bingen,
Auf der Morgenstelle 14, D-72076 T$\ddot{\textrm{u}}$bingen, Germany}
\begin{abstract}
It is a well known fact that Dirac phenomenology of nuclear forces 
predicts the existence of large scalar and vector mean fields in matter. 
To analyse the relativistic self-energy in a model independent 
way, modern high precision nucleon-nucleon ($NN$) potentials are 
mapped on a relativistic operator basis using projection techniques. 
This allows to compare the various potentials at the level of covariant 
amplitudes were a remarkable agreement is found. It allows further to 
calculate the relativistic self-energy in nuclear matter in Hartree-Fock 
approximation. Independent of the choice of 
the nucleon-nucleon interaction 
large scalar and vector mean fields of several hundred MeV 
magnitude are generated at tree level. In the 
framework of chiral EFT these fields are dominantly generated by 
contact terms which occur at next-to-leading order in the chiral 
expansion. Consistent with Dirac phenomenology the corresponding low energy 
constants which generate the large fields are closely connected to the 
spin-orbit interaction in $NN$ scattering. 
The connection to QCD sum rules is discussed as well. 

\end{abstract}
\maketitle

\section{Introduction}
A fundamental question in nuclear physics is the role which 
relativity plays in nuclear systems. The ratio of the Fermi momentum 
over the nucleon mass is about $k_{\rm F}/M \simeq 0.25$ and 
nucleons move with maximally about 1/4 of the velocity of light. This 
implies only moderate corrections from relativistic kinematics in finite 
nuclei. 
Non-relativistic approaches such as,\\e.g., Skyrme-Hartree-Fock and 
relativistic approaches describe finite nuclei equally well. 

However, there exists a fundamental difference between relativistic 
and non-relativistic dynamics: a genuine feature of relativistic 
nuclear dynamics is the appearance of large scalar and vector 
mean fields, each of a magnitude of several hundred MeV. The scalar 
field $\Sigs$ is attractive and the vector field  $\Sigma_\mu $ is 
repulsive. In relativistic mean field (RMF) theory, both, the sign 
and the size 
of the large scalar and vector fields are enforced by the nuclear 
saturation mechanism \cite{sw86}.  
At nuclear saturation density $\rho_0\simeq 0.16~{\rm fm}^{-3}$ 
the empirical fields deduced from RMF fits to finite nuclei are 
of the order of  $\Sigs\simeq -350$ MeV and 
$\Sigo \simeq +300$ MeV \cite{rmf} (In mean field theory only the time-like 
component of $\Sigma_\mu$ contributes in static systems with time-reversal 
symmetry).  

A problem is, however, that these scalar/vector fields are no 
direct observables as, e.g., the nuclear binding energy or the 
nucleon potential. The single particle potential in which the nucleons move 
originates from the cancellation of the two contributions 
$U_{\rm cent} \simeq \Sigo + \Sigs $ and 
is of the order of -50 MeV. Therefore one has no direct experimental 
access to the interpolating scalar/vector fields. 
There exist, however, several 
features in nuclear structure which can be explained naturally
within Dirac phenomenology while models based on non-relativistic dynamics 
have difficulties or, at least, one has
to introduce additional model parameters. The most well 
known feature is the large {\it spin-orbit splitting} in finite nuclei. In  
a relativistic framework 
the strong spin-orbit force appears naturally from the 
coupling to the lower components of the Dirac equation 
where the scalar-vector mean fields add up in the spin-orbit 
potential $U_{\rm S.O.} \propto  (\Sigo -  \Sigs) \simeq 750$ MeV. 
Due to this fact 
RMF theory is able to reproduce the strong spin-orbit 
splitting in spherical nuclei quantitatively 
without the introduction of additional parameters \cite{rmf}. A second 
symmetry, observed more than thirty years ago in single-particle 
levels of spherical nuclei is the so called {\it pseudo-spin symmetry} 
\cite{arima69}. 
While all attempts to understand this symmetry within non-relativistic 
approaches failed, it can naturally be understood within RMF theory as 
has been shown by Ginocchio~\cite{ginocchio97} a few years ago. This symmetry, 
again a consequence of the coupling to the lower components,  
is exact in the limit $\Sigo = -  \Sigs$ and is broken in 
nature by the amount $(\Sigo +  \Sigs)/(\Sigo - \Sigs) $ 
 which is less than 10\%. A third example are the {\it moments of inertia 
in rotating nuclei}. Relativistic dynamics implies that in the 
rotating system a Coriolis term occurs due to the spatial vector 
currents, however, with all couplings already fixed through the 
time-like components~\cite{ring96b}.   

An alternative approach for nuclear matter are {\it ab initio} 
many-body calculations. Based on high precision nucleon-nucleon ($NN$)
interactions 
one treats short-range and many-body correlations explicitly. A typical 
example for a successful many-body approach is Brueckner theory~\cite{gammel}. 
In the relativistic Brueckner approach the nucleon 
inside the medium is dressed by the self-energy $\Sigma$. 
The in-medium T-matrix is obtained from 
the relativistic Bethe-Salpeter (BS) equation and plays the role 
of an effective two-body interaction which contains all short-range 
and many-body correlations of the ladder approximation. 
Solving the BS equation the Pauli principle is respected 
and intermediate scattering states are projected 
out of the Fermi sea. 
The summation of the  T-matrix over the occupied states inside the Fermi sea 
yields finally the self-energy in Hartree-Fock 
approximation~\cite{anastasio83,terhaar87a,brockmann90,gross99,schiller01,dalen04}. 
In contrast to relativistic Dirac-Brueckner-Hartree-Fock (DBHF) 
calculations which came up in the late
80ties non-relativistic BHF theory has already half a century's
history~\cite{gammel}. Despite strong 
efforts invested in the development of improved solution techniques for 
the Bethe-Goldstone (BG) equation, the non-relativistic counterpart of the 
BS equation, it turned out that, although such calculations were able to
describe the nuclear saturation mechanism qualitatively, they failed 
quantitatively. Systematic studies for a large number of nucleon-nucleon
interactions showed that saturation points were always allocated on a
so-called {\it Coester-line} in the $E/A-\rho$ plane which does not
meet the empirical region of saturation. In particular modern 
one-boson-exchange (OBE) potentials 
lead to strong over-binding and to too large saturation densities where 
relativistic calculations work much better \cite{brockmann90,honnef}. 

However, in relativistic approaches the nuclear interaction is 
always described in some sort of a meson exchange picture. The mesons 
represent effective bosonic degrees of freedom which are either directly 
adjusted to the properties of nuclear matter and finite nuclei, as in the 
case of RMF theory, or to vacuum $NN$ scattering. 
Hence it is a fundamental question to 
decide whether the large scalar and vector fields enforced by 
Dirac phenomenology of nuclear systems are an artefact of the
meson exchange picture or whether they reflect a deeper characteristics 
of nature. 

A connection to Quantum-Chromo-Dynamics (QCD) 
as the fundamental theory of strong interactions 
is established by QCD sum rules~\cite{cohen91,drukarev91}. The change  
of the chiral condensates 
$\langle {\bar q}q \rangle, \langle q^\dagger q \rangle $ in matter leads to 
attractive scalar and repulsive vector self-energies which 
are astonishingly close to the empirical values 
derived from RMF fits to the nuclear chart.

It is remarkable that relativistic many-body calculations yield 
again scalar and vector fields which are of the same sign and magnitude 
as obtained from RMF theory or, alternatively, from QCD sum rules. 
Such a coincidence could not have been expected a priori. Moreover, 
 DBHF calculations  \cite{gross99} agree even on a quantitative level
surprisingly well with the QCD based approach of Ref. 
\cite{finelli} where chiral fluctuations from 
the long and intermediate range pion-nucleon dynamics were considered 
on top of the chiral condensates. 

These facts suggest that preconditions for the 
existence of large fields in matter or, 
alternatively, the density dependence of the QCD condensates, must already 
be inherent in the vacuum $NN$ interaction. The 
connection of the nucleon-nucleon force to QCD is given 
by the fact that  the interaction is described by the exchange of the low 
lying mesonic degrees of freedom. The long-range part of the interaction is 
mediated by the one-pion-exchange (OPE) while the scalar 
isoscalar intermediate range
attraction is mainly due to correlated two-pion-exchange. The short-range
part, i.e., the hard core, is dominated by light vector meson exchange, 
i.e., the vector isoscalar $\omega$ meson and the vector isovector $\rho$. 
Modern one-boson-exchange potentials (OBEP) as e.g.\ the Bonn 
potentials~\cite{bonn} are based on the exchange of these mesons and 
provide high precision fits to nucleon-nucleon scattering data. 
Meson-nucleon coupling constants and form factors are empirically fixed 
from the data. Thus OBEPs are the result of relativistic phenomenology 
at the level of the elementary $NN$ interaction. There exist, however, also 
high precision non-relativistic empirical potentials such as the 
Argonne potential \cite{av18} or the Nijmegen 
potentials~\cite{nijmegen}. 
 
A more systematic and direct connection to QCD is provided by 
chiral effective field theory (EFT). Up to now the two-nucleon system 
has been considered at next-to-next-to-next-to-leading order (N$^3$LO) in 
chiral perturbation theory \cite{entem02,entem03,epelbaum05}. 
In such approaches 
the $NN$ potential consists of one-, two- and three-pion exchanges and 
contact interactions which account for the short-range contributions. 
The advantage of such approaches is the systematic expansion of the $NN$ 
interaction in terms of chiral power counting. The expansion is performed 
in powers of $(Q/\Lambda_\chi)^\nu$ where $Q$ is the generic low 
momentum scale given by the nucleon three-momentum, or the four-momenta of 
virtual pions or a pion mass. 
$\Lambda_\chi \simeq 4\pi f_\pi \simeq 1$ GeV is the chiral
symmetry breaking scale which coincides roughly with the Borel mass 
$\Lambda_B$ of the sum rules. In such an expansion the 
low-energy constants (LECs) related to pion-nucleon 
vertices can be fixed from pion-nucleon scattering data \cite{epelbaum05}. 

A better understanding of the common features and the differences of the 
various approaches is essential in order to arrive at a more model independent 
understanding of the $NN$ interaction, in particular since 
all the well established interactions fit $NN$ scattering data with 
approximately the same precision. A direct comparison of relativistic 
phenomenology based on the meson 
exchange picture with chiral EFT and non-relativistic phenomenology is, 
however, difficult 
since the latter two approaches lack of a clear Lorentz structure.  
At low momentum scales the different potentials 
can be mapped on each other using renormalization group methods
\cite{lowk03}. This led recently to 
the construction of a ``model independent'' 
low momentum potential $V_{\rm low~ k}$ by integrating out
the dynamics for momenta above a cut-off scale of about $\Lambda \simeq 2
~{\rm fm}^{-1}$~\cite{lowk03}. It has been argued that beyond this 
scale the short-range part of the interaction, mediated by vector 
meson exchange or point-like counter terms, becomes dominant and 
leads to the deviations of the various approaches.   

Although a breakthrough in some sense, the renormalization group approach 
does not help to clarify the relativistic structure of the potentials 
which is essential e.g.\ in order to generate (or not to generate) large 
scalar/vector mean fields in nuclear matter.  

The present work tries to answer this question. We apply projection 
techniques to map the various potentials on Dirac phenomenology. The 
philosophy behind this approach is based on the fact, that any $NN$ 
interaction, 
independent whether relativistic or non-relativistic, contains a certain 
spin-isospin operator structure.  
By projection techniques this operator basis is 
mapped on the operator basis 
of Dirac phenomenology which is given by the Clifford algebra 
in Dirac space. This allows to identify the different Lorentz 
components of the interaction. Starting from the 
angular-momentum representation of a given $NN$ potential, 
one transforms to plane-wave helicity 
states and finally to Lorentz invariant amplitudes in Dirac 
space~\cite{tjon85a,horowitz87}. Such a transformation is well defined 
in the positive energy sector for on-shell amplitudes and allows 
to compare the $NN$ potentials on the basis of Lorentz invariant 
amplitudes. A remarkable agreement between relativistic and non-relativistic
OBE potentials, non-relativistic phenomenological potentials and EFT 
potentials, respectively, has been found in~\cite{plohl06}. 
\begin{figure}
\begin{center}
\includegraphics[width=0.85\textwidth] {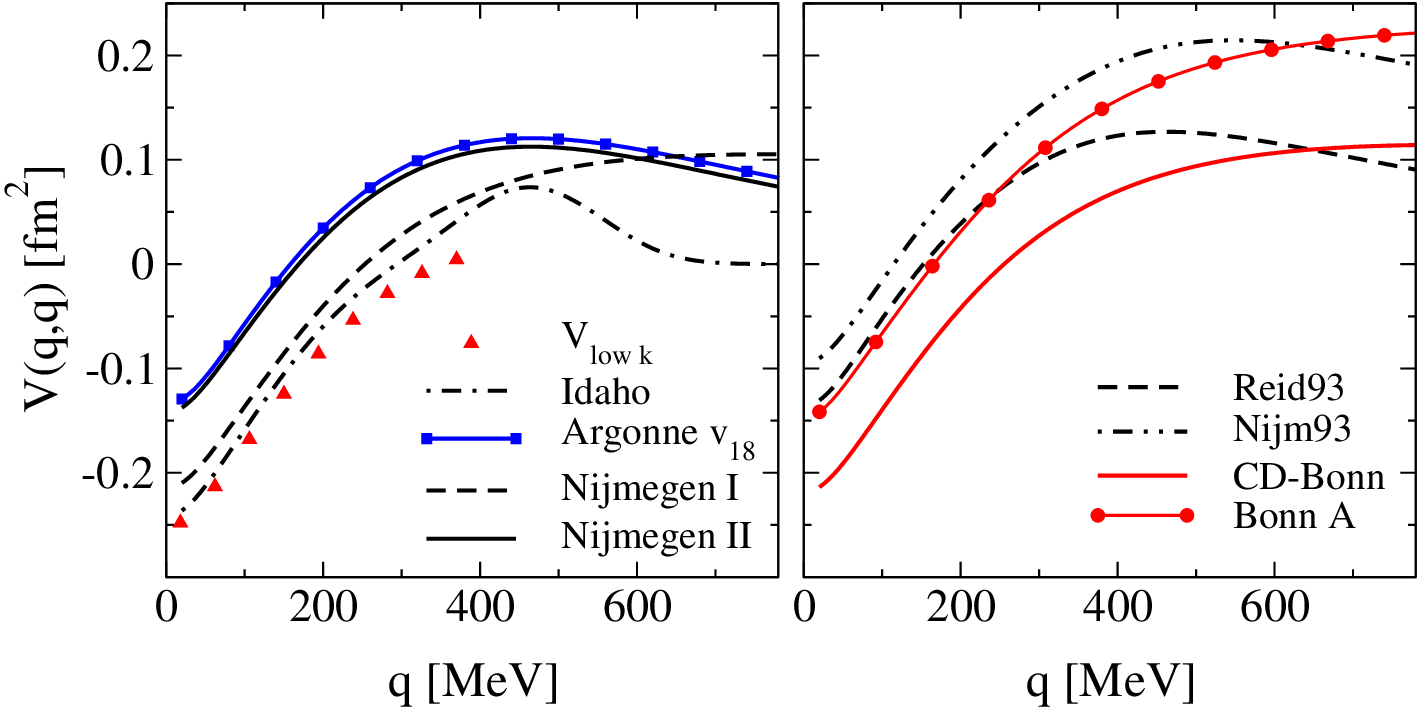}
\caption {(Color online) Diagonal matrix elements $V({\bf q},{\bf q})$ 
in the $^1S_0$ partial wave for different high precision $NN$
potential models.  
\label{partialwaves} }
\end{center}
\end{figure}
This agreement is also 
reflected in the structure of the relativistic self-energy when 
we further calculate the 
mean field in infinite nuclear matter in Hartree-Fock approximation 
at tree-level.  The scalar and vector self-energy components are 
found to be large, i.e., of the order of several hundred MeV~\cite{plohl06}. 
The present work extends the investigations of Ref.~\cite{plohl06}. The 
formalism is outlined in detail and we discuss the density dependence of 
the fields as well as the implications for the nuclear equation-of-state. 
The connection between chiral EFT and QCD sum rules is investigated. 
The present formalism allows a quantitative extraction of the scalar/vector 
fields which are generated by pion dynamics and contact terms at different 
orders in the chiral expansion. 

The paper is organised as follows: in sec. II we discuss the operator 
structure of the various potentials. The transformation onto the 
covariant basis is outlined in sec. III where also the results of this 
analysis namely the Lorentz invariant amplitudes are shown. 
Sec. IV contains the determination of the relativistic mean fields in 
nuclear matter.
In sec. V the structure of the relativistic self-energy fields 
from chiral EFT is discussed, as well as the connection to QCD sum rule
predictions. Finally the self-consistent Hartree-Fock results for 
the equation of state calculated with three different
potentials (Bonn A, Nijm93 and Nijm~I) are discussed in sec. VI.

\section{Operator structure of the $NN$ potentials}
\subsection{OBE potentials}

As typical examples for modern high precision OBEPs we consider 
the Bonn A \cite{machleidt89} and the high-precision, charge-dependent Bonn 
potential (CD-Bonn) \cite{cdbonn}. 
The Bonn potentials are based on the exchange of 
the six non-strange bosons $(\pi, \eta, \rho, \omega, \delta, \sigma)$ 
with masses below 1 GeV. These are the two 
scalar mesons $\sigma$ (isoscalar) and $\delta$  (isovector), 
the two pseudo-scalar mesons  $\pi$ (isovector) 
and $\eta$ (isoscalar), and the two vector mesons  $\omega$ (isoscalar) 
and $\rho$ (isovector). The potentials are derived in the {\it no sea} 
approximation which neglects the coupling to anti-particles. 

The Born scattering matrix 
is given by the sum over the corresponding scalar, pseudo-scalar and 
vector mesons and has the following structure 
\begin{equation}
{\hat V} (q^\prime , q)
=\sum_{\alpha=s,ps,v} {\cal F}^{2}_{\alpha}(q^\prime , q)~ 
\kappa_{\alpha}^{(2)} ~
D_{\alpha}(q^\prime - q) ~\kappa_{\alpha}^{(1)} ~~.
\label{vobe1}
\end{equation}
In the two-nucleon centre-of-mass frame (c.m.) the four-momenta of the 
incoming nucleons are $q_{\mu}^{(1/2)} = (E({\bf q}), \pm {\bf q})$ 
and correspondingly, the four-momenta of the outgoing nucleons are
$q_{\mu}^{\prime (1/2)}= (E({\bf q}^\prime), \pm {\bf q}^\prime)$. The 
initial and final relative c.m. momenta are 
  $q_{\mu} = \frac{1}{2}(q_{\mu}^{(1)}- q_{\mu}^{(2)})$ and 
$q_{\mu}^{\prime }=\frac{1}{2}(q_{\mu}^{\prime (1)}- q_{\mu}^{\prime (2)})$, 
respectively. For on-shell scattering 
$|{\bf q}| =|{\bf q}^\prime|$ with $E({\bf q}) = E({\bf q}^\prime) = 
\sqrt{M^2 + {\bf q}^2}$ the energy-transfer is zero, i.e.,  
$q^{\prime}_\mu - q_\mu = (0,{\bf q}^\prime - {\bf q})$. The matrices 
(\ref{vobe1}) factorise for each meson $\alpha$ into the 
form factors ${\cal F}_{\alpha}$ at each meson-nucleon vertex, the meson propagator $D_\alpha$  
and the meson-nucleon vertices $\kappa_{\alpha}$ themselves. 
In the standard Bonn potentials \cite{machleidt89} the phenomenological 
form factors have the form 
\begin{equation}
{\cal F}_{\alpha}(q^\prime , q) = \left( \frac{\Lambda_{\alpha}^2 - 
m_{\alpha}^2}{\Lambda_{\alpha}^2 + ({\bf q}^\prime - {\bf q})^2}
\right)^{n_\alpha}
\label{obeform}
\end{equation}
where $m_{\alpha}$ is the corresponding meson mass and $\Lambda_{\alpha}$ 
is a cut-off in order to avoid divergences at short distances. 
The meson propagators read 
\begin{eqnarray}
D_{s,ps}(q^\prime - q) = i\frac{1}{(q^\prime - q)^2 - m_{s,ps}^2}\quad, \quad
D_{v}^{\mu\nu}(q^\prime - q) = i\frac{-g^{\mu\nu} +(q^\prime - q)^\mu 
(q^\prime - q)^\nu/m_{v}^2}{(q^\prime - q)^2 - m_{v}^2}
\label{mprop1}
\end{eqnarray}
for scalar and pseudo-scalar mesons $s,ps$ and vector mesons $v$. 
The Dirac structure of the potential is contained in the 
meson-nucleon vertices 
\begin{eqnarray}
\kappa_{s}=\frac{g_{s}}{(2\pi)^2}{\bf{1}},\quad   
  \kappa_{ps}=\frac{g_{ps}}{(2\pi)^2}
\frac{\fmslash{q}^\prime - \fmslash{q}}{2M} i{\gamma^5} 
, \quad 
 \kappa_{v}=\frac{1}{(2\pi)^2}\left( g_v{\gamma^{\mu}}
+ \frac{f_v}{2M} i\sigma^{\mu\nu}\right)~~.
\label{vertex1}
\end{eqnarray}
For the pseudo-scalar mesons $\pi$ and $\eta$ a pseudo-vector coupling 
is used in order to fulfil soft pion theorems. The vertices of the 
isovector bosons $\pi,\delta,\rho$ obtain additional 
${\bf \tau}_2\cdot {\bf \tau}_1$ isospin matrices which are suppressed in 
Eqs.~(\ref{vertex1}). The $\omega$ meson has no tensor coupling, i.e., 
$f_{v}^{(\omega)} =0$. 

The potential, i.e., the OBE Feynman amplitudes are obtained by 
sandwiching ${\hat V}$ between the incoming and outgoing Dirac spinors
\begin{equation}
V ({\bf q}',{\bf q}) 
=\sum_{\alpha=s,ps,v} {\cal F}^{2}_{\alpha}({\bf q}',{\bf q})~ 
D_{\alpha}( {\bf q}^\prime - {\bf q}) ~
\bar{u}_2(-{\bf q}')\kappa_{\alpha}^{(2)} u_2(-{\bf q}) ~
\bar{u}_1({\bf q}')\kappa_{\alpha}^{(1)} u_1({\bf q})~~.
\label{vobe2}
\end{equation}
The relativistic operator structure is thus completely determined by 
the matrix elements of the vertices 
$\kappa_\alpha$. In helicity representation the Dirac spinor basis
is given by
\beqa
u_\lambda ( {\bf q})= \sqrt{\frac{E+ M}{ 2M }} 
\left( 
\begin{array}{c} 1 \\ 
{2\lambda |{\bf q}|}\over{E+M}
\end{array}
\right)
\chi_\lambda~~,
\eeqa
where $\chi_\lambda$ denotes a two-component Pauli spinor with 
$\lambda=\pm {1\over 2}$. The normalisation of the Dirac spinor is 
chosen such that $\bar{u}_\lambda u_\lambda =1$. 

A consequence of the Feynman amplitudes~(\ref{vobe2}) is their general 
non-local structure which distinguishes the field theoretical relativistic 
OBE approach from local non-relativistic potentials. This is even true 
for the relativistic OPE compared to the local, non-relativistic OPE 
(see e.g.\ the discussion in~\cite{machleidt01}). 
However, for on-shell scattering the relativistic amplitudes acquire a 
local structure in the sense that they are functions of ${\bf q}^2$ and 
${\bf q}'-{\bf q}$. In particular for forward and backward scattering,  
i.e., $\theta =0,\pi$, the amplitudes are ``local'' functions of 
${\bf q}^2$ and ${\bf q}$. The non-local structure of the relativistic 
amplitudes becomes evident when going off-shell, e.g.\  in the 
intermediate states in the Bethe-Salpeter equation 
\cite{muether00,machleidt01}. 

The standard Bonn (A,B,C) potentials~\cite{machleidt89} contain 
13 free parameters for coupling constants and cut-off masses and 
two additional parameters if one considers the masses of the 
scalar mesons as 
effective parameters. The matrix elements are calculated 
with the OBNNS code of R. Machleidt~\cite{machleidt93} when Bonn A 
is used and the corresponding CDBONN package of R. Machleidt when 
CD-Bonn is used. 

In contrast to the standard Bonn 
potentials~\cite{machleidt89} the OPE part of the CD-Bonn 
potential~\cite{cdbonn} accounts 
for charge symmetry breaking in  $nn,~pp$ and $np$ scattering 
due to the different pion masses $m_{\pi^0}$ and $m_{\pi^\pm}$. The CD-Bonn 
potential can be referred to as a phenomenological $NN$ potential, since by 
fine-tuning of the 
partial wave fits $\chi^{2}$ per datum is minimised to 1.02, adding up 
to a total of 43 free parameters. 
\subsubsection{Non-relativistic reduction}\label{nrreduction}
The OBE potentials as e.g.\ the Bonn potentials can be reduced to a 
non-relativistic representation by expanding the full 
field-theoretical OBE Feynman amplitudes into a set of 
spin and isospin operators
\begin{equation}
V= \sum_{i} [V_i+V_i'\,\bftau_1\!\cdot\!\bftau_2]\,\,O_i. 
\label{non-rel.V}
\end{equation}
The operators $O_i$ obtained in this low energy expansion, assuming identical particle scattering and charge independence, are defined as
\begin{equation}
  \begin{array}{l}
   O_{1}=1,  \\                                 
   O_{2}=\mbox{\boldmath $\sigma$}_{1}\!\cdot\!
         \mbox{\boldmath $\sigma$}_{2},               \\[0.2cm]
   O_{3}=(\mbox{\boldmath $\sigma$}_{1}\!\cdot\!{\bf k})
         (\mbox{\boldmath $\sigma$}_{2}\!\cdot\!{\bf k})
         \\[0.2cm]
   O_{4}={\textstyle\frac{i}{2}}
        (\mbox{\boldmath $\sigma$}_{1}+\mbox{\boldmath $\sigma$}_{2})
        \cdot{\bf n},\\                         
   O_{5}=(\mbox{\boldmath $\sigma$}_{1}\!\cdot\!{\bf n})
         (\mbox{\boldmath $\sigma$}_{2}\!\cdot\!{\bf n}), \\[0.2cm]
   \end{array}    \label{Pmom}
\end{equation}
where ${\bf k}={\bf q'}-{\bf q}$, ${\bf n}={\bf q}\times{\bf q'}\equiv{\bf P}
\times {\bf k}$ and ${\bf P}=\frac{1}{2}({\bf q}+{\bf q'})$
is the average momentum.
The potential forms $V_i$ are then functions of ${\bf k}$, ${\bf P}$, ${\bf n}$
 and the energy.  
In order to perform a non-relativistic reduction, usually the energy $E$ is
 expanded in ${\bf k}^2$ and ${\bf P}^2$
\begin{equation}
E({\bf q}) = \left(\frac{{\bf k}^2}{4} + {\bf P}^2+M^2\right)^{\frac{1}{2}} 
\simeq M+ \frac{{\bf k}^2}{8M}+\frac{{\bf P}^2}{2M}.
\end{equation}
and terms to leading order in ${\bf k}^2/M^2$ and ${\bf P}^2/M^2$ are taken into account. The meson propagators $D_{\alpha}(k^2)$ 
given in Eq.~(\ref{mprop1}) are approximated by their static form 
$(-1)/({\bf k}^2+m^2)$. The equivalent to Eq.~(\ref{non-rel.V}) in 
configuration space is given by
\begin{equation}
  \begin{array}{l}
   O_{1}=1,                                           \\[0.2cm]
   O_{2}=\mbox{\boldmath $\sigma$}_{1}\!\cdot\!
         \mbox{\boldmath $\sigma$}_{2},               \\[0.2cm]
   O_{3}=S_{12}=3(\mbox{\boldmath $\sigma$}_{1}\!\cdot\!\hat{\bf r})
                 (\mbox{\boldmath $\sigma$}_{2}\!\cdot\!\hat{\bf r})
                -\mbox{\boldmath $\sigma$}_{1}\!\cdot\!
                  \mbox{\boldmath $\sigma$}_{2},     \\[0.2cm]
   O_{4}={\bf L\cdot S},                              \\[0.2cm]
   O_{5}=Q_{12}={\textstyle\frac{1}{2}}
         [(\mbox{\boldmath $\sigma$}_{1}\!\cdot\!{\bf L})
          (\mbox{\boldmath $\sigma$}_{2}\!\cdot\!{\bf L})+
          (\mbox{\boldmath $\sigma$}_{2}\!\cdot\!{\bf L})
          (\mbox{\boldmath $\sigma$}_{1}\!\cdot\!{\bf L})]. \\
   \end{array}     \label{Pconfig}
\end{equation}
These operators are the well known central,
spin-spin, tensor, spin-orbit and quadratic spin-orbit operators, respectively.
The total angular momentum is denoted by ${\bf L}={\bf r} \times{\bf P}$ and 
the total spin ${\bf S}=\frac{1}{2}(\mbox{\boldmath $\sigma$}_{1}+\mbox{\boldmath $\sigma$}_{2})$.

\subsection{Non-relativistic potentials}
\subsubsection{Meson-theoretical potentials}
We consider the modern Nijmegen soft-core potential Nijm93~\cite{nijmI_II} as 
the first example of a non-relativistic meson-theoretical potential. It is an updated version of the Nijm78~\cite{nijm78} potential, where the low energy $NN$ 
interaction is based on Regge-pole theory leading to the well known OBE forces.
The contributions considered in this model are the pseudo-scalar mesons 
$\pi$, $\eta$, $\eta'$, the vector mesons $\rho$, $\phi$, $\omega$ and the 
scalar mesons $\delta$, $S^*$, $\epsilon$ and the Pomeron $P$ and the $J=0$ 
tensor contributions, leading all in all to a number of 13 free parameters.
Since it is constructed from approximate OBE amplitudes it is based on 
the operator structure given in Eq.~(\ref{Pmom}) plus an additional operator 
$O_6=\frac{1}{2}(\mbox{\boldmath $\sigma$}_{1}-\mbox{\boldmath $\sigma$}_{2})\cdot{\bf L}$ accounting for charge independence breaking which is new compared 
to the older version Nijm78. 
Exponential form factors are used. This potential gives a $\chi^2$ per datum 
of 1.87, which is comparable to similar OBE potentials like the standard Bonn 
potentials. 
\subsubsection{Phenomenological potentials}
Another class of non-relativistic $NN$ potentials are the so called high 
quality potentials where $\chi^2/N_{data}\approx 1.0$. 
Here we study the Nijmegen 
potentials Nijm I, Nijm II and Reid93~\cite{nijmI_II}. 
The Nijm I and Nijm II potentials are both based on the Nijm78 potential. In 
the Nijm I potential some nonlocal terms in the central force are kept whereas
in the Nijm II potential all nonlocal terms are removed. Although based on the
meson-theoretical Nijm78 potential these 
potentials are often referred to as purely phenomenological models, since the 
parameters are adjusted separately in each partial wave leading to a total 
of 41 parameters. At very short distances, both potentials are regularised 
by an exponential form factor.

The Nijmegen soft-core Reid93~\cite{nijmI_II} potential is a phenomenological 
potential and is therefore based on a completely different approach. 
In the meson-theoretical Nijmegen potential Nijm93 the potential forms $V_i$ 
are the same for all partial waves, whereas in the Reid93 potentials every 
partial wave is parametrised separately by a 
convenient choice of combinations of central, tensor and spin-orbit functions 
(local Yukawas of multiples of the pion mass) and the related operators, 
i.e., the operators $O_1$ to $O_4$ from Eq.~(\ref{Pconfig}). 
It is regularised by a dipole form factor and has 50 phenomenological 
parameters giving all in all a $\chi^2/N_{data}= 1.03$.  
All the Nijmegen potentials contain the proper charge dependent 
OPE accounting 
for charge symmetry breaking in  $nn,~pp$ and $np$ scattering 
due to different pion masses $m_{\pi^0}, m_{\pi^\pm}$.

The same holds for the 
Argonne potential $v_{18}$ \cite{av18}, also an example for a widely 
used modern high precision phenomenological $NN$ potential. It is given by 
the sum of an electromagnetic 
(EM) part, the proper OPE, and a phenomenological intermediate- and short-range
 part unrestricted by a meson-theoretical picture:
\begin{equation}
   V = V^{EM} + V^{\pi} + V^{R} \ .
\label{argonne1}
\end{equation}
The EM interaction is the same as that used in the Nijmegen
partial-wave analysis. Short-range terms and
finite-size effects are taken into account as well~\cite{av18}.

The strong interaction part $V^{\pi} + V^{R}$ can be written in a form 
like given in Eq.~(\ref{non-rel.V}) in configuration space, where 
the Argonne $v_{18}$ potential is not constructed by approximating the 
field-theoretical OBE amplitudes (except for the OPE), but by assuming a 
very general two-body potential constrained by certain symmetries. The 
potential forms $V_i$ parametrising the intermediate and short-range part
are mostly local Woods-Saxon functions.

The local two-body operators are the same charge independent 
ones used in the Argonne $v_{14}$ potential 
\begin{eqnarray}
   O_i &=& 1, {\sigma}_{1}\!\cdot\!{\sigma}_{2},
S_{12}, {\bf L}\!\cdot\!{\bf S},L^{2},
L^{2}({\sigma}_{1}\!\cdot\!{\sigma}_{2}),({\bf L}\!\cdot\!{\bf S})^{2}.
\end{eqnarray}
Due to isovector exchange these operators have to be multiplied by
the isospin matrices ${\bf \tau}_1\cdot {\bf \tau}_2$ which than adds up to 14 
operators. Additionally, four operators accounting for charge independence 
breaking are introduced
\begin{equation}
   O_{i=15,18}  =  T_{12}, \,
        ({\sigma}_{1}\!\cdot\!{\sigma}_{2})
        T_{12},\, S_{12}T_{12},\, (\tau_{z1}+\tau_{z2})\ ,
\end{equation}
where $T_{12}=3\tau_{z1}\tau_{z2}-{\tau}_{1}\!\cdot\!{\tau}_{2}$,
is the isotensor operator, defined analogously to the spin tensor 
$S_{12}$ operator.

Thus the operator structure is more general than that imposed by a 
non-relativistic, local OBE picture, in particular for the intermediate and 
 short distance part.  
In total,  Argonne $v_{18}$ contains 40 adjustable parameters and 
gives a $\chi^{2}$ per datum
of 1.09 for 4301 $pp$ and $np$ data in the range 0--350 MeV \cite{av18}. 
The code used to calculate the potential matrix elements of the 
Argonne $v_{18}$ model in momentum space was provided by H. Muether and 
T. Frick.
\subsection{Low-Energy potentials}

\subsubsection{EFT potentials}

Following the concept originally proposed by Weinberg \cite{Wei90} 
there has been substantial progress in recent time in order to 
derive quantitative $NN$ potentials from chiral effective field
theory. As already mentioned, the chiral expansion is performed 
in powers of $(Q/\Lambda_\chi)^\nu$ where $\nu=0$ corresponds 
to leading order (LO),  $\nu=2$ to next-to-leading order (NLO),  
$\nu=3$ to next-to-next-to-leading (N$^2$LO) and finally $\nu=4$ to 
next-to-next-to-next-to-leading 
order (N$^3$LO). It turned out that for a quantitative description of 
$NN$ scattering data one has to go up to N$^3$LO \cite{entem02,entem03,epelbaum05} 
in the chiral expansion for the two-nucleon problem.  
 N$^2$LO contributions were still found to be very large compared 
to NLO. This implies that $2\pi$ (and $3\pi$) contributions
have to be included up to order four. 
The effective chiral Lagrangian can be written as
\begin{equation}
{\cal L}_{\rm eff} 
=
{\cal L}_{\pi\pi}^{(2)} 
+
{\cal L}_{\pi N}^{(1)} 
+
{\cal L}_{\pi N}^{(2)} 
+
{\cal L}_{\pi N}^{(3)} 
+ \ldots ,
\label{chlag}
\end{equation}
where the superscript refers to the number of derivatives or 
pion mass insertions (chiral dimension)
and the ellipsis stands for terms of chiral order four or higher.
The corresponding chiral $NN$ potential is then defined by
\begin{equation}
V({\bf q}~', {\bf q}) \equiv
\left\{ \begin{array}{c}
\mbox{ sum of irreducible}\\
\mbox{\boldmath $\pi + 2\pi$ contributions}
\end{array} \right\} 
+ \mbox{ contacts} \, .
\label{eq_pot1}
\end{equation}
The $2\pi$ exchange contributions to the $NN$ interaction
at order four have been derived by 
Kaiser~\cite{Kai01}. Recently, quantitative $NN$ potentials including 
contact terms at order four were derived by Entem and 
Machleidt, the so-called Idaho potential \cite{entem02,entem03},  
and by Epelbaum, Gl\"ockle and Meissner \cite{epelbaum05}. 

For the present comparison we apply the Idaho potential \cite{entem03}. 
The operator structure of the momentum-space $NN$ amplitude has the general
form given in Eq.~(\ref{non-rel.V}) 
with the operators $O_i$ from Eq.~(\ref{Pmom}). 
The potential forms $V_i$ ($i=C,S,T,LS,\sigma L$) can be expressed 
as functions of $|({\bf q}^\prime - {\bf q})|$ and $ |{\bf k}|$.

The Idaho potential is regularised by an exponential 
cut-off 
\begin{equation}
V({\bf q}^\prime  ,{{\bf q}}) \longmapsto 
V({\bf q}^\prime  ,{{\bf q}})
\;\mbox{\boldmath $e$}^{-(q'/\Lambda)^{2n}}
\;\mbox{\boldmath $e$}^{-(q/\Lambda)^{2n}}
\label{cut1}
\end{equation}
where $\Lambda = 0.5$ GeV in all partial waves. 
This does not affect the chiral order of 
the potential, but introduces contributions beyond that order.
The total number of free model parameters in the N$^3$LO potential 
is 29~\cite{entem03}. 

For the evaluation of the matrix elements we applied the N$^3$LO program 
package provided by D. Entem and R. Machleidt. 
\subsubsection{Renormalization Group approach to $NN$ interaction}
Recently, another approach has been proposed to arrive at a better 
model independent understanding of the $NN$ interaction~\cite{lowk03}. In this
approach a low-momentum potential $V_{\rm low~k}$ is derived from a given 
realistic $NN$ potential  by separating the low-momentum part, i.e., by 
integrating out the high-momentum modes, 
and using renormalization group (RG) methods to
 evolve the $NN$ potential models from the full Hilbert space to the 
low momentum subspace.
At a cutoff of $\Lambda =2.1$ fm$^{-1}$ all the various $NN$ potential models
were found to collapse to a model-independent effective 
interaction $V_{\rm low~ k}$.

Since elastic $NN$ scattering data constrains the $NN$ interaction only 
up to a momentum scale of about 400 MeV, which corresponds to the pion 
threshold, modern high precision potentials differ essentially 
in the treatment of  the short-range part, 
as depicted in Fig.~\ref{partialwaves}.
The philosophy behind the RG approach 
is to replace the unresolved short distance structure by something simpler, 
e.g.\ contact terms, without distorting low-energy observables.

\section{Transformation to a covariant operator basis}
%

%
\subsection{Covariant operators in Dirac space}
Any two-body amplitude can be represented covariantly by 
Dirac operators and Lorentz invariant amplitudes. A detailed 
discussion of the general structure of relativistic two-body 
amplitudes can be found in Refs.~\cite{tjon85a,tjon85b}. 
However, a relativistic treatment invokes 
automatically the excitation of anti-nucleons. Nucleon-nucleon scattering, 
in both, the non-relativistic approaches discussed above but also in the 
framework of the standard OBE potentials is restricted to the positive 
energy sector and neglects the coupling to anti-nucleons. As a consequence 
one has to work in a subspace of the full Dirac space which leads to 
on-shell ambiguities which require some care. 

The inclusion of negative energy excitations with 4 states for 
each spinor yields altogether $4^4=256$ types of two-body matrix elements 
with respect to their
spinor structure. Symmetry arguments reduce these to 44 for on-shell particles
\cite{tjon85b}. If one takes only the subspace of 
positive energy solutions into account
this leads to $2^4=16$ two-body matrix elements. Considering in addition only
on-shell matrix elements the number of independent matrix elements 
can be further reduced by symmetry arguments down to 5. 
Thus, all on-shell two-body matrix elements
can be expanded into five Lorentz invariants. These five 
invariants are not unique since the Dirac matrices involve always 
also negative energy states. Therefore
a decomposition of the one-body $NN$ potential into 
a Lorentz scalar and a Lorentz vector contribution depends to some 
part on the choice of these five Lorentz invariants. 

A natural choice of a set of five
linearly independent covariant operators to represent a $4\times 4$ 
Dirac matrix are the scalar, vector, tensor, axial-vector and 
pseudo-scalar Fermi covariants 
\beq
 {\rm S}=1\otimes1,\quad {\rm V}=\gamma^{\mu}\otimes\gamma_{\mu},\quad
  {\rm T}=\sigma^{\mu\nu}\otimes \sigma_{\mu\nu},\quad
{\rm P}=\gamma_5\otimes\gamma_5,\quad {\rm A}=\gamma_5\gamma^{\mu}\otimes \gamma_5\gamma_{\mu}. 
\label{cov1}
\eeq
Since one works with physical,
i.e., antisymmetrized matrix elements, one has to realize that 
the Fierz transformation ${\cal F}$~\cite{tjon85a} couples direct 
and exchange covariants which mixes the different Lorentz structures 

\parbox{11cm}{
\begin{displaymath}
\pmatrix{
{\tilde {\rm S}} \cr {\tilde {\rm V}} \cr {\tilde {\rm T}} \cr {\tilde {\rm
    A}} \cr {\tilde {\rm P}}  \cr } = {\displaystyle \frac{\displaystyle 1}{\displaystyle 4}}
\pmatrix{
1 & 1 & \frac{1}{2} & -1 & 1 \cr 
4 & -2 & 0 & -2 & -4 \cr
12 & 0 & -2 & 0 & 12 \cr
-4 & -2 & 0 & -2 & 4  \cr
1 & -1 & \frac{1}{2} & 1 & 1 \cr
}
\pmatrix{
{\rm S} \cr {\rm V} \cr { {\rm T}} \cr { {\rm A}} \cr {{\rm P}} \cr
}
\end{displaymath}
} \hfill
\parbox{0.5cm}
{\begin{eqnarray}\label{fierz} \end{eqnarray}}

The covariants on the left hand side of
Eq.~(\ref{fierz}) are the interchanged Fermi covariants defined in
Ref.~\cite{tjon85a} as
\beq
{\rm {\tilde S}={\tilde S}S,\quad {\tilde V}={\tilde S}V, \quad {\tilde
  T}={\tilde S}T, \quad {\tilde A}={\tilde S}A, \quad {\tilde
  P}={\tilde S}P, }
\label{cov2}
\eeq
where the operator ${\tilde {\rm  S}}$ exchanges the Dirac indices of
particles 1 and 2, i.e., ${\tilde {\rm
    S}}u(1)_{\sigma}u(2)_{\tau}=u(1)_{\tau}u(2)_{\sigma}$. Therefore the
direct covariants $\Gamma_m$ with $m=\{{\rm S,V,T,P,A}\}$ can be
expressed in terms of the exchange covariants ${\tilde \Gamma_m}$ with 
$m=\{{\rm {\tilde S},{\tilde V},{\tilde T},{\tilde P},{\tilde A}}\}$.

In contrast to the $NN$ potentials where the pion-nucleon coupling is 
given by a pseudo-vector vertex, the set (\ref{cov1},\ref{cov2})  
contains the pseudo-scalar covariant P. This suggests to replace P in 
Eqs.~(\ref{cov1},\ref{cov2}) by the corresponding pseudo-vector covariant
\beq
{\rm PV}=\frac{\fmslash{q}^\prime -\fmslash{q}}{2M} \gamma_5\otimes
\frac{\fmslash{q}^\prime -\fmslash{q}}{2M}\gamma_5~~.
\eeq
This leads to an on-shell equivalence 
since the matrix elements of the pseudo-vector and the
pseudo-scalar matrix operators are identical in the case of on-shell
scattering between positive energy states: 
\beq
{\bar u}({\bf q}^\prime  )\frac{\fmslash{q}^\prime -\fmslash{q}}{2M}\gamma_5
u({\bf q})
={\bar u}({\bf q}^\prime   )\gamma_5u({\bf q}) ~~.
\label{pv1}
\eeq
On the other hand the PV vertex suppresses a coupling 
to antiparticles since the overlap matrix elements vanish 
for on-shell scattering 
\beq 
{\bar v}({\bf q}^\prime  ) \frac{\fmslash{q}^\prime -\fmslash{q}}{2M}\gamma_5 u({\bf q})=0 ~~ .
\label{pv2}
\eeq
In order to identify the PV contributions clearly in the antisymmetrized 
amplitudes - note that due to the Fierz transformation (\ref{fierz}) all 
operators are coupled - one can switch to a set of covariants originally 
proposed 
by Tjon and Wallace \cite{tjon85b}. Based on the following operator identities 
\beq
\frac{1}{2}({\rm T + {\tilde T})=S+{\tilde S}+P+{\tilde P}}
\eeq
\beq
{\rm V+{\tilde V}=S+{\tilde S}-P-{\tilde P}}
\eeq
one finds that the following set of covariants 
\beq
\Gamma_m = \{ {\rm S},{\rm {\tilde S}},({\rm A}-  {\rm {\tilde A}}),{\rm PV} ,
{\rm {\widetilde {PV}}}\}
\label{cov3}
\eeq
provides a set of Dirac operators for the positive energy sector
\cite{tjon85b} which completely separates the direct and exchange 
$pv$ contributions from the remaining operator structure. This has the
advantage that the OPE exchange which is dominant at 
low energies is decoupled from the remaining amplitudes and gives 
only a contribution to the ${\rm {\widetilde {PV}}}$ operator. 
In the following we will refer to the set of covariants in Eq.~(\ref{cov3}) as 
the $pseudo-vector$ representation and that of Eq.~(\ref{cov1}) as 
the $pseudo-scalar$ representation. Note that on-shell matrix elements of 
${\rm PV,{\widetilde {PV}}}$ in~(\ref{cov3}) are equivalent to those where 
the pseudo-vector covariants are replaced by ${\rm P,{\tilde P}}$. 

The on-shell equivalence does not affect physical observables 
which are built on complete matrix 
elements as e.g.\ 
the single particle potential $U$  
\beq
U({\bf k})_{\rm s.p.} \propto \sum_{{\bf q}} ~
\langle {\bar u}({\bf k}){\bar u}({\bf q}) |
{\hat V} ({\bf k},{\bf q}) |u({\bf k}) u({\bf q}) 
-  u({\bf q}) u({\bf k})  \rangle 
\label{usp1}
\eeq
but it leads to uncertainties in operators which are, like 
the self-energy $\Sigma$, based on traces over only one particle.  
As discussed in \cite{fuchs98}, a pseudo-vector $\pi N$ coupling 
leads to the $pseudo-vector$ representation (\ref{cov3}) 
as the most natural choice of the relativistic 
operator basis.

\subsection{Projection onto the covariant operators}

In this section the technique is described necessary to project the Born 
amplitudes from an angular-momentum basis onto 
the covariant basis, given by Eqs.~(\ref{cov1}) or~(\ref{cov3}).  
The procedure is standard and runs over the following steps 
\beqa 
 |LSJ\rangle \rightarrow {\rm partial~ wave~ helicity~ states} 
\rightarrow {\rm plane~ wave~ helicity~ states}\rightarrow {\rm covariant~ basis}. 
\nonumber
\eeqa
The first two transformation can be found 
in Refs.~\cite{erkelenz74,machleidt89}. The last step depends on the choice 
of the covariant operator basis, see e.g.\ ~\cite{horowitz87,gross99}.  
Here we sketch the essential steps briefly.

Independent of the various models, the amplitudes are
determined normally in the $|LSJM\rangle$-representation and can be denoted
as $V^{JS}_{L',L}(\vecq^\prime,\vecq)$. In case of on-shell
scattering (${\qabs}=|\vecq^\prime |$),  due to time-reversal 
invariance and spin and parity conservation, only
five of sixteen possible matrix elements
are linearly independent for a fixed total angular momentum $J$ (spin singlet
and triplet states). 
By inversion of Eq.~(3.32) in~\cite{erkelenz74} these five partial
wave amplitudes  are transformed from the $ |LSJM\rangle$-representation 
into the partial wave helicity
representation $ | JM \lambda_1 \lambda_2 \rangle$ and are then decoupled
via inversion of Eq.~(3.28) from Ref.~\cite{erkelenz74}. 
Since we deal with two-nucleon states which are two-fermion states, we
have to evaluate the fully antisymmetrized matrix elements by
restoring the total isospin ${\rm I}=0,1$ via the
standard selection rule
\beq
(-1)^{L+S+{\rm I}}=-1.
\label{select}
\eeq
The five plane wave helicity matrix elements are then obtained by a
summation over the total angular momentum $J$
\beq
\langle\lambda_1'\lambda_2'\,\vecq^\prime\left|V^{\rm I}\right|\lambda_1\lambda_2\,\vecq\rangle=\sum_J\left(\frac{2J+1}{4\pi}\right
)d^J_{\lambda\,\lambda'}(\theta)\langle
\lambda_1'\lambda_2' | V^{J,{\rm I}}(\vecq^\prime,\vecq) | \lambda_1\lambda_2 \rangle ~~.
\label{sum}
\eeq
Here $\theta$ denotes the scattering angle between ${\bf q}^\prime$ and
${\bf q}$ while $\lambda=\lambda_1-\lambda_2$ and
$\lambda'=\lambda_1'-\lambda_2'$ denote the in- and outgoing 
helicity states. The reduced rotation matrices
$d^J_{\lambda\,\lambda'}(\theta)$ are those defined by Rose~\cite{rose57}.

These plane wave helicity matrix elements can now be projected onto a set of
five covariant amplitudes in Dirac space. A set of
five linearly independent covariants is sufficient for such a
representation since on-shell we deal with five matrix elements
independent of the chosen representation. Using the 
covariants of Eq.~(\ref{cov1}) (the 'pseudo-scalar choice') the on-shell
potential matrix elements for definite isospin $I$ can be represented 
covariantly as~\cite{horowitz87}
\begin{eqnarray}
{\hat V}^{\rm I}(|{\bf{q}}|,\theta)  & = &  
F_{\rm S}^{\rm I}(|{\bf{q}}|,\theta)\,{\rm
 S}+F_{\rm V}^{\rm I}(|{\bf{q}}|,\theta)\,{\rm V}+F_{\rm
 T}^{\rm I}(|{\bf{q}}|,\theta)\,{\rm T}\nonumber  \\
 & & + F_{\rm P}^{\rm I}(|{\bf{q}}|,\theta)\,{\rm P}+F_{\rm
 A}^{\rm I}(|{\bf{q}}|,\theta)\,{\rm A}~~.
\label{ps-sc}
\end{eqnarray}
The Lorentz invariant amplitudes
$F_m^{\rm I}(|{\bf{q}}|,\theta)$ with $m=\{{\rm S,V,T,P,A}\}$ 
from Eq.~(\ref{ps-sc}) depend 
only on  the relative c.m. momentum $|{\bf{q}}|$
and the scattering angle $\theta$ and are related 
to the plane wave helicity states defined in Eq.~(\ref{sum}) by
\beq
\langle\lambda_1'\lambda_2'\,\vecq^\prime |V^{\rm I}  |\lambda_1\lambda_2\,\vecq\rangle
= \sum_m
\langle\lambda_1'\lambda_2'\,\vecq^\prime |\Gamma_m
|\lambda_1\lambda_2\,\vecq\rangle F_m^{\rm I}(|{\bf{q}}|,\theta)~~.
\label{expansion}
\eeq
The indices (1) and (2) refer to particle one and two. 
Eq.~(\ref{expansion}) is a matrix relation between the five
independent plane wave helicity amplitudes $V_i^{\rm I}$ (where
$i=\{\lambda_1',\lambda_2',  \lambda_1,\lambda_2\}=1,...,5$ denotes
five of sixteen possible amplitudes) and the five unknown
covariant amplitudes $F_m^{\rm I}(|{\bf{q}}|,\theta)$. For fixed values of
the variables
$(\qabs=|{\bf{q}}^\prime|,\theta)$  this equation can be written in a
more compact form 
\beq
V^{\rm I}_i=\frac{1}{M^2}\sum_m C_{im}F_m^{\rm I}.
\label{matrix_eq} 
\eeq
The covariant amplitudes $F_m^{\rm I}$ are obtained by matrix inversion of
Eq.~(\ref{matrix_eq}) which corresponds to Eq.~(3.23) of
Ref.~\cite{horowitz87}.

Eq.~(\ref{matrix_eq}) has to be inverted for two
scattering angles, i.e., for $\theta=0$ for the direct and 
$\theta=\pi$ for the exchange part of the interaction. These two scattering 
angles are required for the Hartree-Fock potential. 
Details of the inversion of Eq.~(\ref{matrix_eq}), as
well as the treatment of kinematical singularities of the matrix $C_{im}$ 
occurring at $\theta=0$ and $\theta=\pi$ are given in appendix C of
Ref.~\cite{horowitz87} where Eq.~(\ref{matrix_eq}) is 
explicitly given for $\theta=0$ and $\theta=\pi$ (Eqs. (C10,11)). 
Following Ref.~\cite{horowitz87} we calculate the real part 
of the five Lorentz invariant
amplitudes $F^{{\rm    I}=0,1}_m(\qabs,\theta=0,\pi)$ 
for the direct and exchange case in both, the isospin
singlet and triplet channels. When derived from physical 
partial wave amplitudes which are already antisymmetrized 
according to the selection rule~(\ref{select}), the exchange 
amplitudes $F_m(\qabs,\pi)$ contain redundant information.

Since we are restricted to the subspace of 
positive energy states, the choice of a set of five 
linearly independent covariants suffers from on-shell ambiguities, 
as discussed above. Thus the set of covariants~(\ref{cov3}) is a more 
appropriate choice~\cite{gross99}. 
In this representation the scattering matrix 
reads~\cite{tjon85b,gross99}
\begin{eqnarray}
{\hat V}^{\rm I}(|{\bf{q}}|,\theta)  & = & 
g_{\rm S}^{\rm I}(|{\bf{q}}|,\theta)\,{\rm S}
-g_{\rm {\tilde S}}^{\rm I}(|{\bf{q}}|,\theta)\,{\rm {\tilde S}} 
+g_{\rm A}^{\rm I}(|{\bf{q}}|,\theta)\,({\rm A-{\tilde A}}) \nonumber \\
 &  & + g_{\rm PV}^{\rm I}(|{\bf{q}}|,\theta)\,{\rm PV} -g_{\widetilde {\rm PV}}^{\rm I}(|{\bf{q}}|,\theta)\, {\widetilde {\rm PV}}~~.
\label{pv_full}
\end{eqnarray}
The new amplitudes $g_m^{\rm I}$ are related to the Lorentz invariant
amplitudes $F_m^{\rm I}$ from Eq.~(\ref{ps-sc}) by the linear
transformation

\parbox{11cm}{
 \begin{displaymath}
\pmatrix{
g_{\rm S}^{\rm I} \cr g^{\rm I}_{\tilde {\rm S}} \cr g_{\rm A}^{\rm I} \cr g_{\rm PV}^{\rm I} \cr g^{\rm I}_{\widetilde {\rm PV}} \cr
}
= {\displaystyle \frac{\displaystyle 1}{\displaystyle 4}} 
\pmatrix{
  4 & -2 & -8 & 0 & -2 \cr 0 & -6 & -16 & 0 & 2 \cr 0 & -2 & 0 & 0 & -2
  \cr 
0 & 2 & -8 & 4 & 2  \cr 0 & 6 & -16 & 0 & -2 \cr
}
\pmatrix{
F^{\rm I}_{\rm S} \cr F^{\rm I}_{\rm V} \cr { F^{\rm I}_{\rm T}} \cr { F^{\rm I}_{\rm P}} \cr {F^{\rm I}_{\rm A}} \cr
}
\end{displaymath}} \hfill 
\parbox{0.5cm}{\begin{eqnarray}\label{fierz2} \end{eqnarray}}

As mentioned before, the representation of the potential given in
 Eq.~(\ref{pv_full})
 has the advantage that the OPE contribution to the amplitudes is completely 
decoupled from the rest of the interaction. The OPE contributes only in the 
pseudo-vector exchange amplitude  $g^{\rm OPE}_{\widetilde {\rm PV}}$ and 
vanishes in all other amplitudes $g_{\rm S}^{\rm OPE} = 
g^{\rm OPE}_{\tilde {\rm S}} = g_{\rm A}^{\rm OPE} = g_{\rm PV}^{\rm OPE}
=0$. Thus one avoids that the low momentum behaviour of these four 
amplitudes is to large extent dominated by OPE exchange contributions 
which are present in all five amplitudes $F_m^{\rm I}$ 
from Eq.~(\ref{ps-sc}) due to the Fierz transformation. In order to compare 
the various potentials at the level of covariant amplitudes the 
pseudo-vector representation is therefore the most efficient and 
transparent one.

\subsection{Covariant amplitudes}
\begin{figure}
\begin{center}
\includegraphics[width=0.85\textwidth] {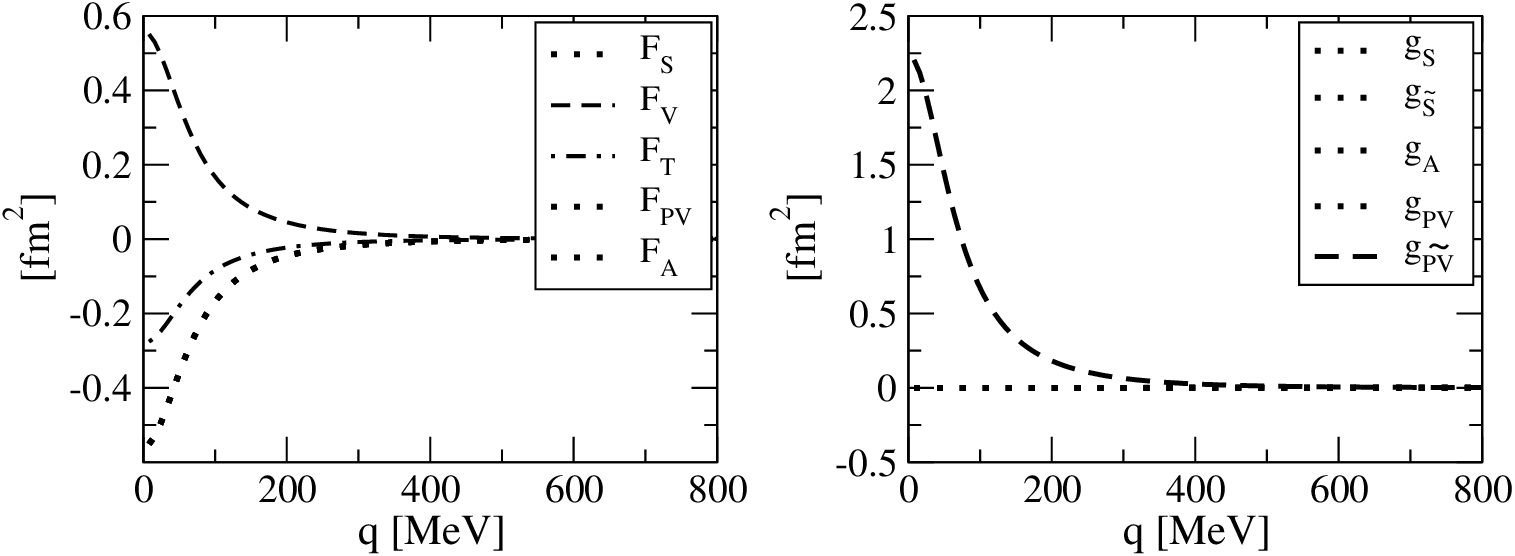}
\caption {Covariant amplitudes for the single OPE for the different 
choices of the relativistic operator basis, i.e., the 
 pseudo-scalar representation $F_m(\qabs,\theta=0)$ (left) and 
the pseudo-vector representation $g_m(\qabs,\theta=0)$ (right).  
\label{pion_ps_fullpv} }
\end{center}
\end{figure}
In order to demonstrate the dependence of the relativistic amplitudes on 
the choice of the operator basis we consider in Fig.~\ref{pion_ps_fullpv} 
first the single OPE. The figure shows the corresponding 
amplitudes $F_m$ of the pseudo-scalar representation (\ref{cov1}) and 
the $g_m$ amplitudes of  pseudo-vector representation (\ref{cov3}), both 
for the OPE part of the Bonn A potential. Since we are dealing with 
antisymmetrized amplitudes it is sufficient to consider the 
direct Lorentz invariants $F_m(\qabs,\theta=0)$ and 
$g_m(\qabs,\theta=0)$ at scattering angle $\theta =0$. 
As the starting point the OPE is given in the $|LSJ\rangle$ basis and 
antisymmetrization is ensured by the selection rule (\ref{select}). 
The figure shows the isospin averaged amplitudes defined as 
\beq
F_m(\qabs,0):=\frac{1}{2}\left[F_m^{{\rm
    I}=0}(\qabs,0)+3F_m^{{\rm I}=1}(\qabs,0)\right]
\eeq
and correspondingly for $g_m$. It is evident that in the 
 pseudo-scalar representation all amplitudes $F_m$ have large non-vanishing 
contributions from OPE due to the mixing of direct and exchange 
contributions described by the Fierz transformation (\ref{fierz}).  
Moreover, as discussed above the on-shell equivalence for 
the pseudo-scalar covariant ${\rm P}$ and the 
pseudo-vector covariant ${\rm \widetilde{PV}}$ in ~(\ref{ps-sc})  
leads to identical Lorentz
invariant amplitudes $F_{\rm PS}=F_{\rm PV} \equiv F_{\rm P}$ ~\cite{gross99}. 
The  pseudo-vector representation (\ref{cov3}), on the other hand, has the 
advantage that it decouples the OPE contribution from the remaining 
amplitudes, i.e., the OPE gives a non-zero contribution only in the 
$g_{\widetilde {\rm PV}}$ amplitude while the others are zero. For the 
single pion exchange $g_{\widetilde {\rm PV}}$ is now easy to interpret: it 
is just the pion propagator (\ref{mprop1}) times the pion-nucleon 
form factor (\ref{obeform}). 

\begin{figure}
\epsfig{file=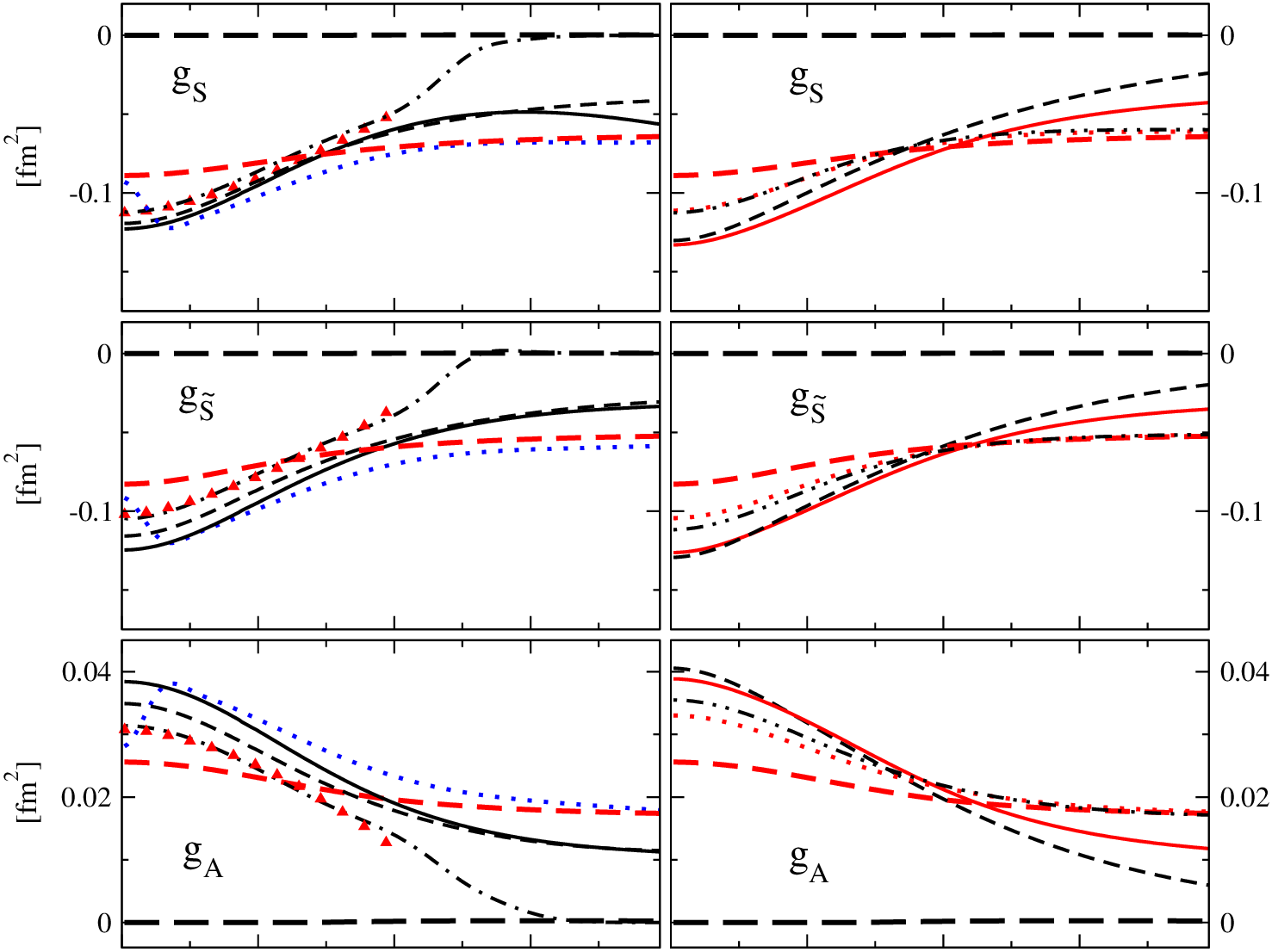,scale=0.5,angle=0}
\epsfig{file=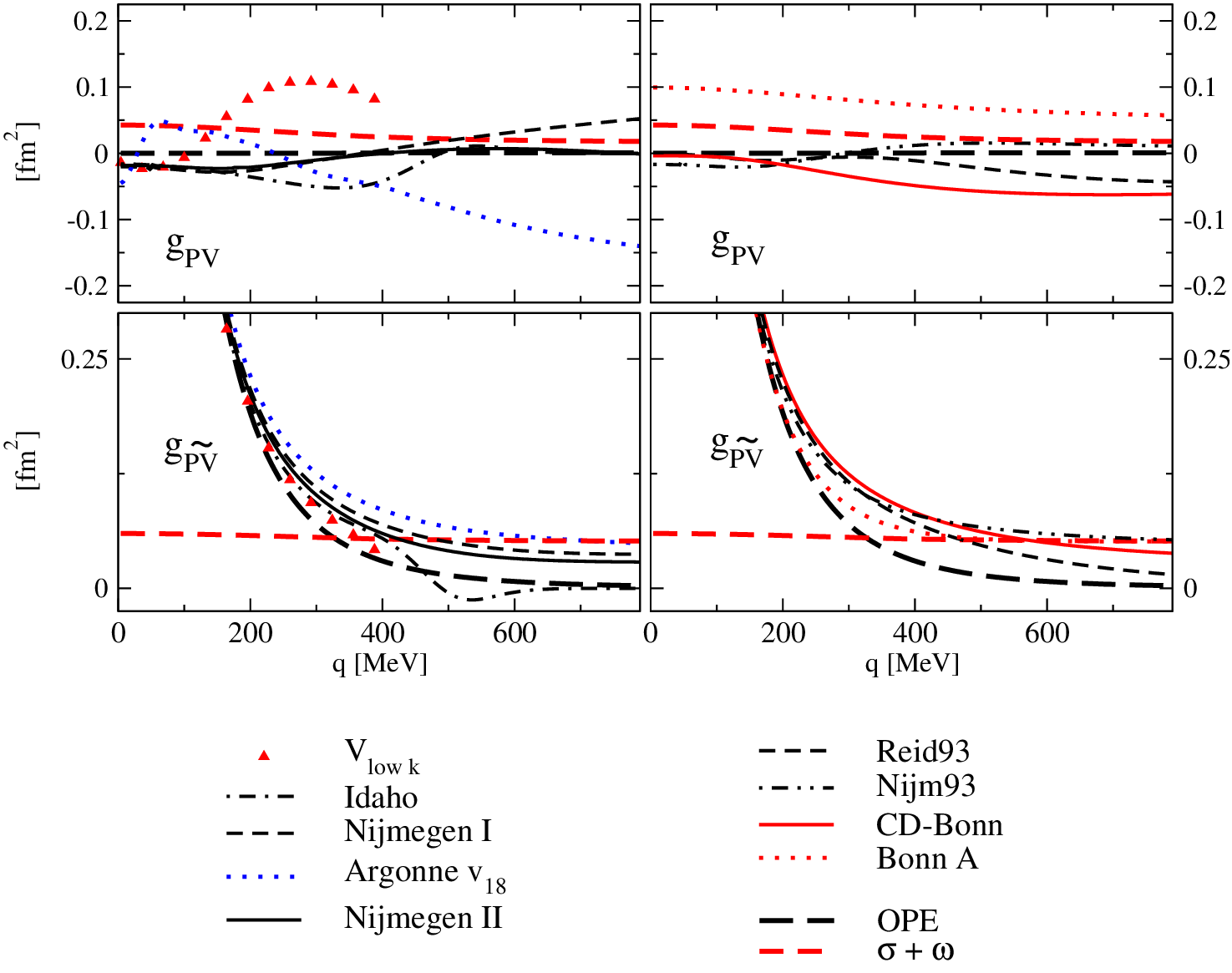,scale=0.5,angle=0}
\caption {(Color online) Isospin-averaged Lorentz invariant
  amplitudes $g^D_m(|{\bf q}|,\theta=0)$ for the 
different $NN$ potentials after projection on the Dirac 
operator structure. The pseudo-vector representation of the 
relativistic operator basis is used. As a reference the 
amplitudes from solely OPE and from $\sigma+\omega$ 
exchange, both with Bonn A parameters, are shown. 
\label{F_pv_d_Itotal} }
\end{figure}
When the various $NN$ potentials are compared, this is done most efficiently 
in the pseudo-vector representation. All potentials contain an OPE of 
similar strength which dominates at small momenta. The  
pseudo-vector representation decouples the OPE contribution from the 
remaining amplitudes $g_m \neq g_{\widetilde {\rm PV}}$ and allows thus 
a more transparent investigation of the short and intermediate range 
parts of the potentials which are actually the interesting ones.   
Fig.~\ref{F_pv_d_Itotal} shows the isospin-averaged amplitudes 
 $g^D_m(|{\bf p}|,\theta=0)$ for Bonn A, CD-Bonn, Argonne $v_{18}$, Nijm93,  
Nijmegen I and II, Reid93, the effective low momentum 
interaction $V_{\rm low~k}$  
and the chiral Idaho potential. The amplitudes are obtained going through 
the transformation scheme discussed above. Partial waves are taken into
account up to $J=90$ (Bonn A, CD-Bonn, Idaho), $J=9$ (Argonne $v_{18}$, 
Nijmegen I/II, Nijm93, Reid93) and $J=6$ ($V_{\rm low~k}$).

The amplitudes determined from the complete 
$NN$ potentials are no more as easy to interpret as for a single 
meson exchange where they represent essentially the propagators times the 
form factors. This is also true for the full OBE 
since the contributions from the various mesons are coupled 
through their exchange parts. Since  these amplitudes are not 
very transparent quantities, Fig.~\ref{F_pv_d_Itotal} includes 
as a reference in 
addition the contributions from only OPE and from only $\sigma$ and 
$\omega$ exchange, both taken from Bonn A.

Several features can now be seen from Fig.~\ref{F_pv_d_Itotal}: First of 
all the four amplitudes  $ g_{\rm S},~g_{\tilde {\rm S}},~ g_{\rm A}$     
and  $ g_{\widetilde {\rm PV}}$ are very close for the OBEPs 
 Bonn A, CD-Bonn and Nijm93 and the phenomenological 
non-relativistic Argonne $v_{18}$ and Nijmegen I/II potentials. Only 
at very small $|{\bf q}|$  Argonne $v_{18}$ shows a 
deviating structure. The direct 
pseudo-vector amplitude  $g_{\rm PV}$ falls somewhat out of systematics. 
This amplitude is, however, of minor 
importance since it does not contribute to the Hartree-Fock 
self-energy (\ref{sigHF1}-\ref{sigHF3}) and to the single particle 
potential.

The dominance of the OPE at low $|{\bf q}|$ is reflected in the pseudo-vector 
exchange amplitude $ g_{\widetilde {\rm PV}}$ which is at small $|{\bf q}|$ 
almost two orders of magnitude larger than the other amplitudes. 
In the OBEPs the high momentum part of the interaction, on the other hand, 
 is dominated by heavy meson exchange  
and the corresponding amplitudes 
$ g_{\rm S},~g_{\tilde {\rm S}},~ g_{\rm A}$  approach the 
$\sigma+\omega$ exchange result. Deviations from the 
$\sigma+\omega$ amplitudes, e.g.\ due to exchange of isovector mesons 
$\rho$ and $\delta$ in the OBEPs  are moderate at large $|{\bf q}|$. 
These deviations are more pronounced at small $|{\bf q}|$. 

The remarkable agreement between the OBE amplitudes and those derived 
from  the non-relativistic Argonne $v_{18}$ potential demonstrates two things: 
first of all, it  means that for on-shell scattering  the Argonne $v_{18}$ can 
be mapped on the relativistic operator structure where the local 
phenomenological functions $V_i$, Eq.~(\ref{non-rel.V}), play 
the same role as the meson propagators plus corresponding 
form factors in the meson exchange picture. Secondly, the effective 
treatment of the short-distance physics in Argonne $v_{18}$ is 
very similar to that in the OBE potentials Bonn A, CD-Bonn and Nijm93. 
This fact can be estimated from Fig.~\ref{partialwaves} 
where the $^1S_0$ partial wave
amplitudes are close as well. On the other hand the softer character of the 
Reid93 and also the Nijmegen I and II potentials is reflected clearly in the 
stronger deviation from the $\sigma+\omega$ amplitudes at large $|{\bf q}|$. 

Finally we are turning to the effective 
low momentum potentials $V_{\rm low~k}$ and the 
chiral Idaho  N$^3$LO potential. 
$V_{\rm low~k}$ is only shown up to the intrinsic 
cut-off of 400 MeV. In this momentum range the amplitudes fall 
practically on top of those from the Idaho N$^3$LO potential.   
At low $|{\bf q}|$ the amplitudes derived from Idaho N$^3$LO and $V_{\rm low~k}$ 
behave qualitatively and quantitatively 
like the previous ones, i.e., they are very close to Bonn A, 
CD-Bonn and Argonne $v_{18}$. We conclude that also 
the effective low momentum potentials can be mapped on a relativistic operator 
structure. For the  Idaho N$^3$LO potential which 
is also based on the operator structure given in Eq.~(\ref{Pmom}), 
the functions $V_i$ and $V_i'$ in 
combination with the corresponding operators, derived from fourth 
order $2\pi$ exchange plus 
contact terms, lead to a structure which is similar to that 
imposed by the OBE picture. 
However, clear deviations appear in the cut-off region 
between 400 and 500 MeV. The short-range interactions are strongly 
suppressed by the exponential cut-off form factors 
and as a consequence the Idaho approaches rapidly the OPE result for 
momenta above 400 MeV.

\section{Self-energy in nuclear matter}\label{selfenergyinmatter}
With the covariant amplitudes at hand, one is able to determine 
the relativistic mean field in nuclear matter  
with its scalar and vector components. To do so, 
we calculate the relativistic self-energy $\Sigma$ in Hartree-Fock 
approximation at {\it tree level}. We are thereby not aiming for a realistic
description of nuclear matter saturation properties which would require 
a self-consistent scheme. Moreover,  short-range correlations require 
to base such calculations on the in-medium T-matrix rather than the 
bare potential $V$. This leads to the relativistic 
Dirac-Brueckner-Hartree-Fock scheme which has been proven to 
describe nuclear saturation with quantitatively satisfying accuracy 
\cite{brockmann90,gross99,schiller01,dalen04}. The self-consistent iteration 
of the self-energy in combination with the Dyson equation for the 
in-medium nucleon propagator and the Bethe-Salpeter equation for the in-medium 
T-matrix leads to self-energy components which are qualitatively of 
similar magnitude than the tree level results, as will be seen later on.

 The self-energy is determined by the summation
 of the interaction of a nucleon with four-momentum $k$ with all
 nucleons inside the Fermi sea in Hartree-Fock approximation
\beq
\Sigma_{\alpha\beta}(k,k_F)=-i\int {d^4q\over {(2\pi)^4}}~
G^D_{\tau\sigma}(q)~
[V(\pabs,0)_{\alpha\sigma;\beta\tau}-V(\pabs,\pi)_{\alpha\sigma;\tau\beta}]~~.
\eeq
Since we work with fully antisymmetrized matrix elements which 
contain already the direct (Hartree) and exchange (Fock) 
contributions, it is sufficient to evaluate the 
Hartree integral for the self-energy 
\beq
\Sigma_{\alpha\beta}(k,k_F)=-i\int {d^4q\over {(2\pi)^4}}~
G^D_{\tau\sigma}(q)~
[V^A(\pabs,0)_{\alpha\sigma;\beta\tau}]~~.
\label{propagator}
\eeq
$G^D(q)$ is the Dirac propagator describing the on-shell propagation
of a nucleon with momentum $q$ inside the Fermi sea in 
the nuclear matter rest frame
\beq
G^D(q)=[\fmslash{q}+M]2\pi i\delta(q^2-M^2)\Theta(q_0)\Theta(k_F-\qabs)~~.
\eeq 
The $\Theta$ functions account for the fact that only positive 
energies are considered. 
Here, ${\bf k}$, taken along the z-axis, is the single particle 
momentum of the incoming nucleon in
the nuclear matter rest frame. The relative momentum in the two-nucleon 
c.m. frame where the matrix elements $V$ are evaluated, is given by
$\pabs =\sqrt {s/4-M^2}$, where $s=(E({\bf k})+E({\bf q}))^2-({\bf
  k+q})^2$ is the total energy of the two nucleons.

Using the pseudo-vector representation for the on-shell 
matrix elements $V$, Eq.~(\ref{pv_full}), the self-energy operator reads
\beqa
\Sigma_{\alpha \beta}(k,k_F) & = &  \int \frac{d^3\vecq}{(2 \pi)^3}  
\frac{\Theta (k_F-\qabs)}{4E({\bf q})}
\left\{ (\fmslash{k}_{\alpha \beta}-\fmslash{q}_{\alpha \beta}) 
\frac {2q_{\mu}(k^{\mu}-q^{\mu})} {4M^2} g_{\widetilde {\rm PV}}  \right. 
 \nonumber \\ 
& & + m 1_{\alpha \beta}
  \left[ 4g_{\rm S} - g_{\tilde {\rm S}}+ 4 g_{\rm A} -\frac
    {(k^{\mu}-q^{\mu})^2} {4M^2} g_{\widetilde {\rm PV}} \right] \nonumber \\ 
& & \left. + \fmslash{q}_{\alpha \beta} \left[- g_{\tilde {\rm S}}+ 2 g_{\rm A}-\frac
    {(k^{\mu}-q^{\mu})^2} {4M^2} g_{\widetilde {\rm PV}}\right] \right\}~~. 
\label{subsec:PS;eq:v}
\eeqa
Translational and rotational invariance, hermiticity, parity
conservation, and time reversal invariance determine the Dirac
structure of the self-energy~\cite{horowitz87}. In the nuclear matter
rest frame the self-energy can be written as 
\beqa
\Sigma(k,\kf)= \Sigs (k,\kf) -\gamma_0 \, \Sigo (k,\kf) + 
\bfgamma  \cdot \textbf{k} \,\Sigv (k,\kf).
\label{subsec:SM;eq:self1}
\eeqa
Note that the sign convention for the vector field 
$\Sigma = \Sigs -\gamma_\mu \Sigma^\mu$ with 
$\Sigma_\mu = (\Sigo,\textbf{k}\Sigv)$ in Eq.~(\ref{subsec:SM;eq:self1}) 
is that used standardly in DBHF~\cite{terhaar87a,gross99,honnef}. It 
differs from that used standardly in QHD 
($\Sigma = \Sigs +\gamma_\mu \Sigma^\mu$) and also that of Eqs.~(\ref{sum1}) 
and~(\ref{sum2}).  The self-energy components are Lorentz scalar
functions depending on the Lorentz invariants $k^2$, $k
\cdot  j$ and $j^2$, where $j_{\mu}$ denotes the four-vector baryon
current. In nuclear matter at rest the time-like component 
is just the baryon density and spatial components of the current 
vanish, i.e., $j_\mu = (\rho_B, {\bf 0})$. 
Hence, the Lorentz invariants can be expressed in terms of
$k_0$, $|\veck|$ and $\kf$, where $\kf$ denotes the Fermi momentum.
The components of the self-energy are computed by taking the
respective traces in the Dirac space~\cite{horowitz87,sehn97}
\beq
\Sigs = \frac{1}{4} tr \left[ \Sigma \right],\quad 
\Sigo = \frac{-1}{4} tr \left[ \gamma_0 \, \Sigma \right], \quad 
\Sigv =  \frac{-1}{4|\veck|^2 } 
tr \left[{\bfgamma}\cdot \veck \, \Sigma \right]. 
\label{subsec:SM;eq:trace}
\eeq
In doing so, the Lorentz components  of the 
self-energy operator~(\ref{subsec:PS;eq:v}) are given by
\beqa
\Sigs(k,k_F)  & = & \frac{1}{4} \int \frac{d^3\vecq}{(2
  \pi)^3} \Theta (k_F-\qabs) \frac{M}{E({\bf q})}  \left[ 4g_{\rm S} -
  g_{\tilde {\rm S}}+ 4 g_{\rm A} -\frac {(k^{\mu}-q^{\mu})^2} {4M^2}
  g_{\widetilde {\rm PV}} \right]~, 
\label{sigHF1}\\
& & \phantom{kk}\nonumber \\
\Sigo(k,k_F)  & = &  \frac{1}{4} \int \frac{d^3\vecq}{(2 \pi)^3}  \Theta (k_F-\qabs)  \left[ g_{\tilde {\rm S}}- 2 g_{\rm A}+ \frac{E({\bf k})}{E({\bf q})} \frac {(k^{\mu}-q^{\mu})^2} {4M^2} g_{\widetilde {\rm PV}}\right]
\label{sigHF2}
\eeqa
and
\beqa
\Sigv(k,k_F)  & = & \frac{1}{4} \int \frac{d^3\vecq}{(2
  \pi)^3} \Theta (k_F-\qabs) \frac{\veck\cdot\vecq}{\kabs ^2E({\bf
    q})}  \left[ g_{\tilde {\rm S}}-2 g_{\rm A} + \frac{k_z}{q_z} \frac{(k^{\mu}-q^{\mu})^2} {4M^2}
  g_{\widetilde {\rm PV}} \right]~.
\label{sigHF3}
\eeqa
\begin{figure}
\begin{center}
\includegraphics[width=0.75\textwidth] {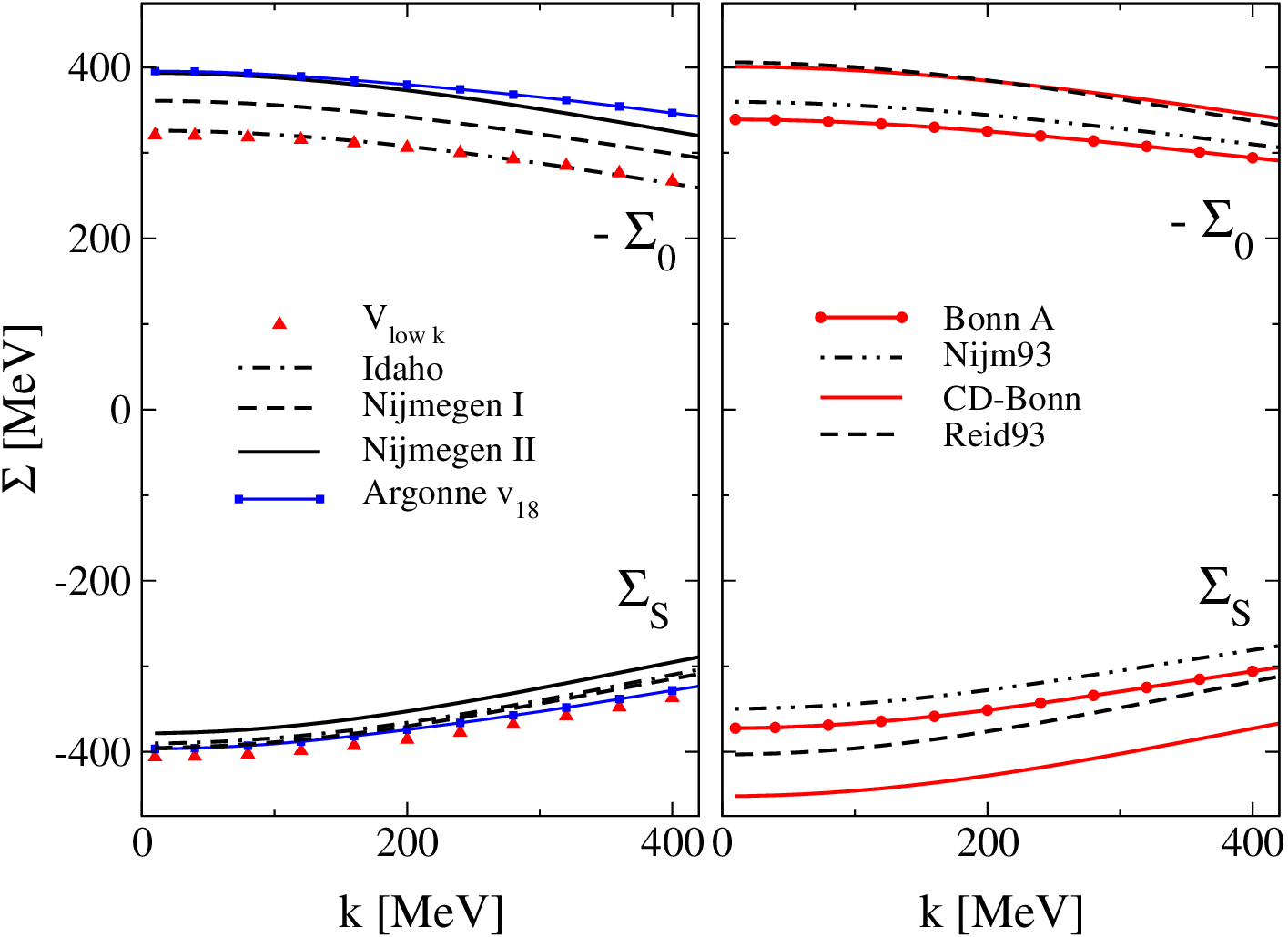}
\caption {(Color online) Tree level scalar and vector self-energy components 
in nuclear matter at $\kf =1.35~{\rm fm}^{-1}$ obtained with 
different $NN$ interaction models. 
\label{fig_sigma1} }
\end{center}
\end{figure}
In Fig.~\ref{fig_sigma1} the tree level scalar and vector self-energy 
components in nuclear matter 
are shown obtained with the various $NN$ potentials 
at nuclear saturation density with Fermi momentum 
$\kf =1.35~{\rm fm}^{-1}$ which corresponds to a density of 
$\rho=0.166~{\rm fm}^{-3}$. As a remarkable result, all potentials 
yield scalar and vector mean fields $\Sigs$ and   $\Sigo$ of comparable 
strength: a large and attractive scalar field  $\Sigs \simeq -(450\div 400)$
MeV and a repulsive vector field of $-\Sigo \simeq +(350\div 400)$ MeV. These 
values are comparable to those derived from RMF phenomenologically and also 
from QCD sum rules. Also the explicit momentum dependence of the self-energy 
is similar for the various potentials. The Idaho mean fields follow 
the other approaches at low $k$ but show a stronger decrease above 
$k\simeq 2~{\rm fm}^{-1}$ which reflects again the influence of the 
cut-off parameter. 
\begin{figure}
\begin{center}
\includegraphics[width=0.75\textwidth] {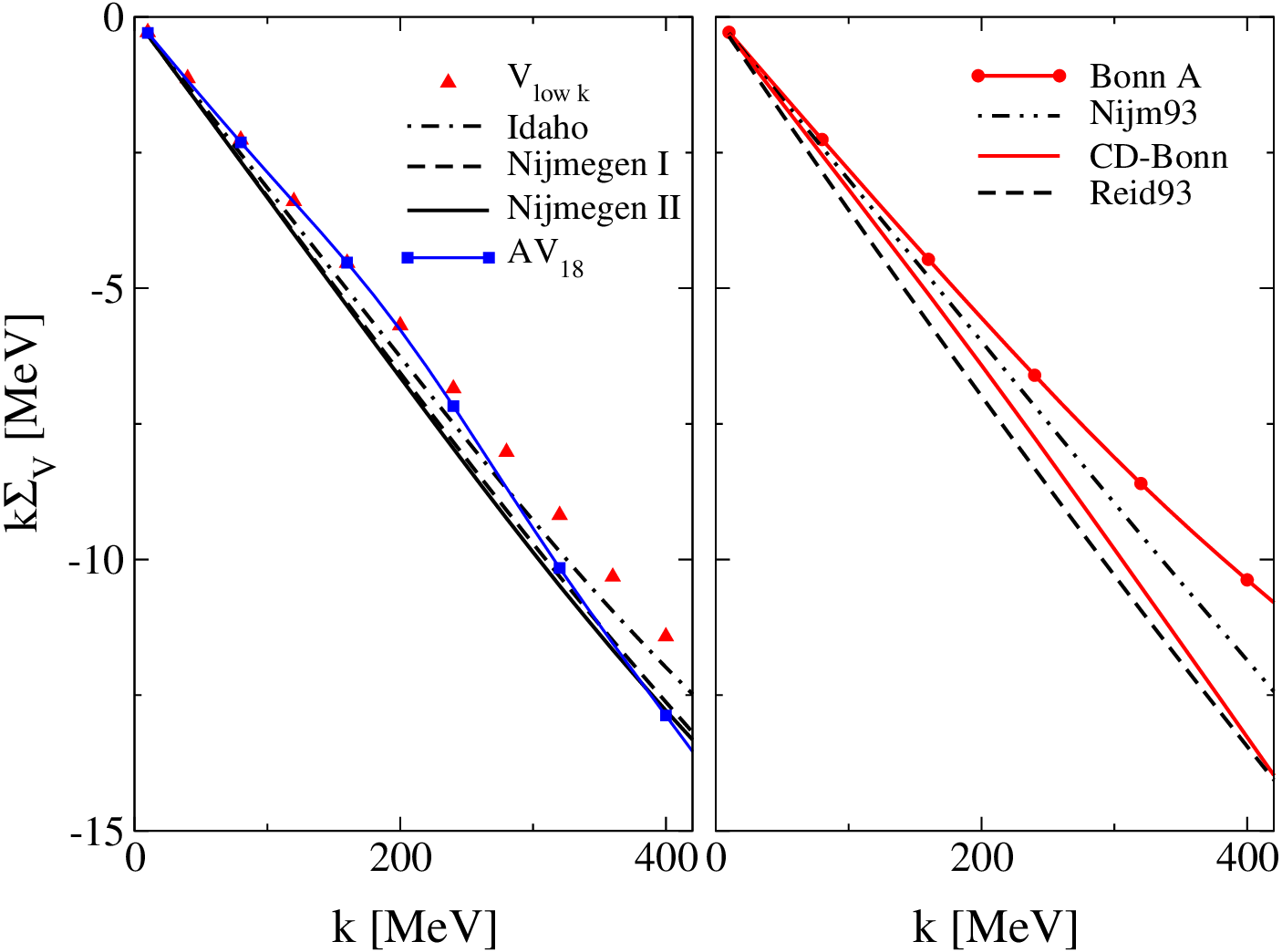}
\caption {(Color online) Tree level spatial vector self-energy component 
${\bf k}\Sigv$  
in nuclear matter at $\kf =1.35~{\rm fm}^{-1}$ for the various potentials. 
\label{fig_sigma2} }
\end{center}
\end{figure}
Fig.~\ref{fig_sigma2} shows the spatial component of the vector self-energy 
${\bf k}\Sigv$, Eq.~(\ref{sigHF3}). 
Also here the various potentials agree quite well. 
As known from self-consistent DBHF calculations~\cite{terhaar87a,gross99}, 
the spatial vector self-energy is a moderate correction to the large 
scalar and time-like vector components $\Sigs$ and $\Sigo$. This is found 
to be also the case  at tree level where 
${\bf k}\Sigv$ is about one order of magnitude smaller than the other 
two components. The spatial self-energy originates exclusively from 
exchange contributions, i.e., the Fock term, and vanishes e.g.\ in the 
mean field approximation of RMF theory. 

\begin{figure}
\begin{center}
\includegraphics[width=0.75\textwidth] {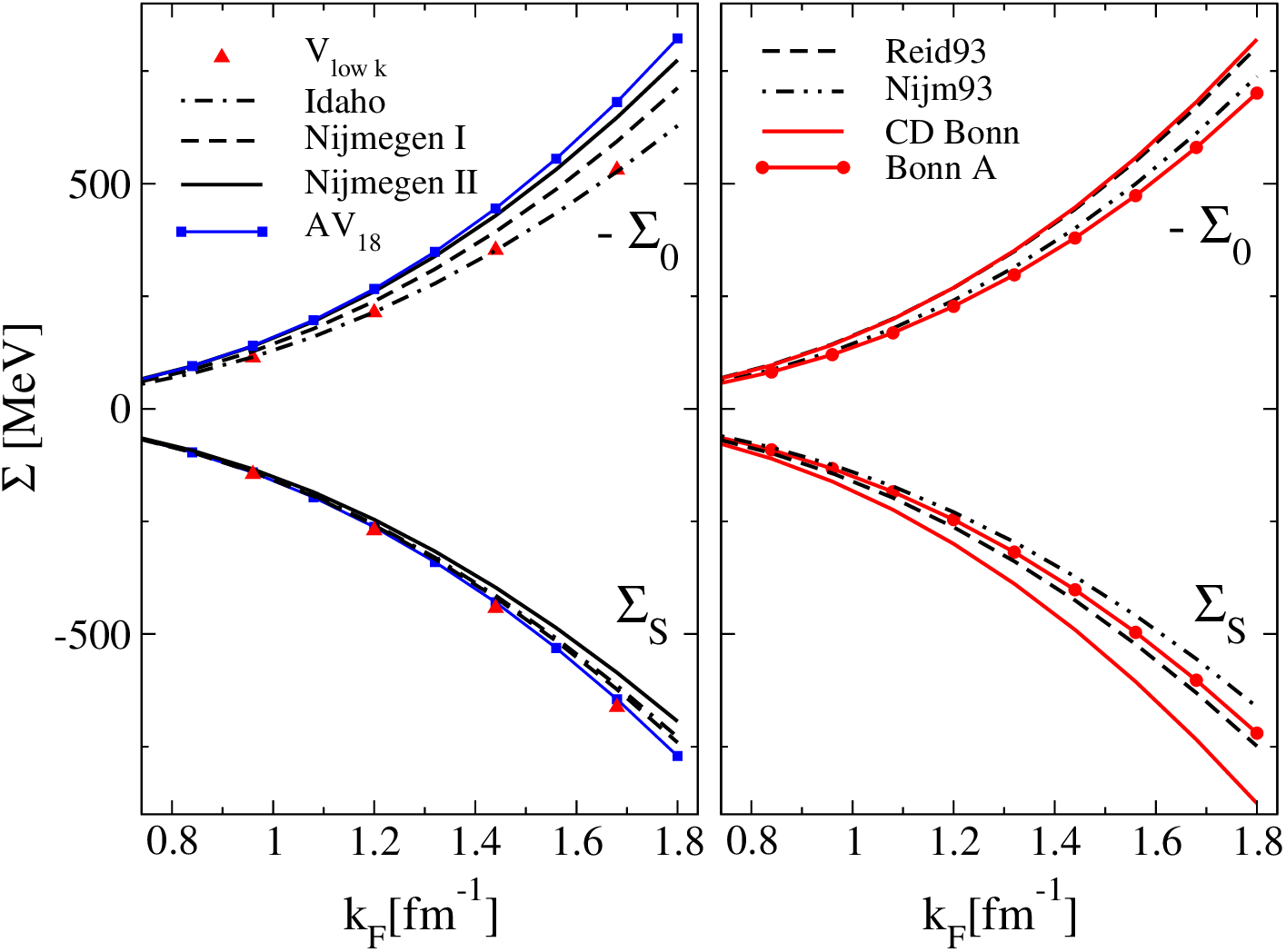}
\caption {(Color online) Density dependence of the tree level scalar and 
vector self-energy components in nuclear matter obtained with 
the various potentials. 
\label{sigma3} }
\end{center}
\end{figure}

Fig.~\ref{sigma3} displays the density dependence of the fields, 
evaluated at momentum $k=\kf$. At moderate densities the different potentials 
yield scalar and vector fields which are rather close in magnitude. At 
higher densities the results start to split up which reflects again 
the different treatment of short distance physics in the various interactions. 
Only the two low momentum interactions Idaho N$^3$LO and $V_{\rm low~k}$ 
lie practically on top of each other. In this context we want to stress 
again that these results are obtained in lowest order in density. Hence, the 
results are only 'realistic' in the low density limit but not at 
higher densities since short-range correlations are missing. 

\begin{figure}
\begin{center}
\includegraphics[width=0.75\textwidth] {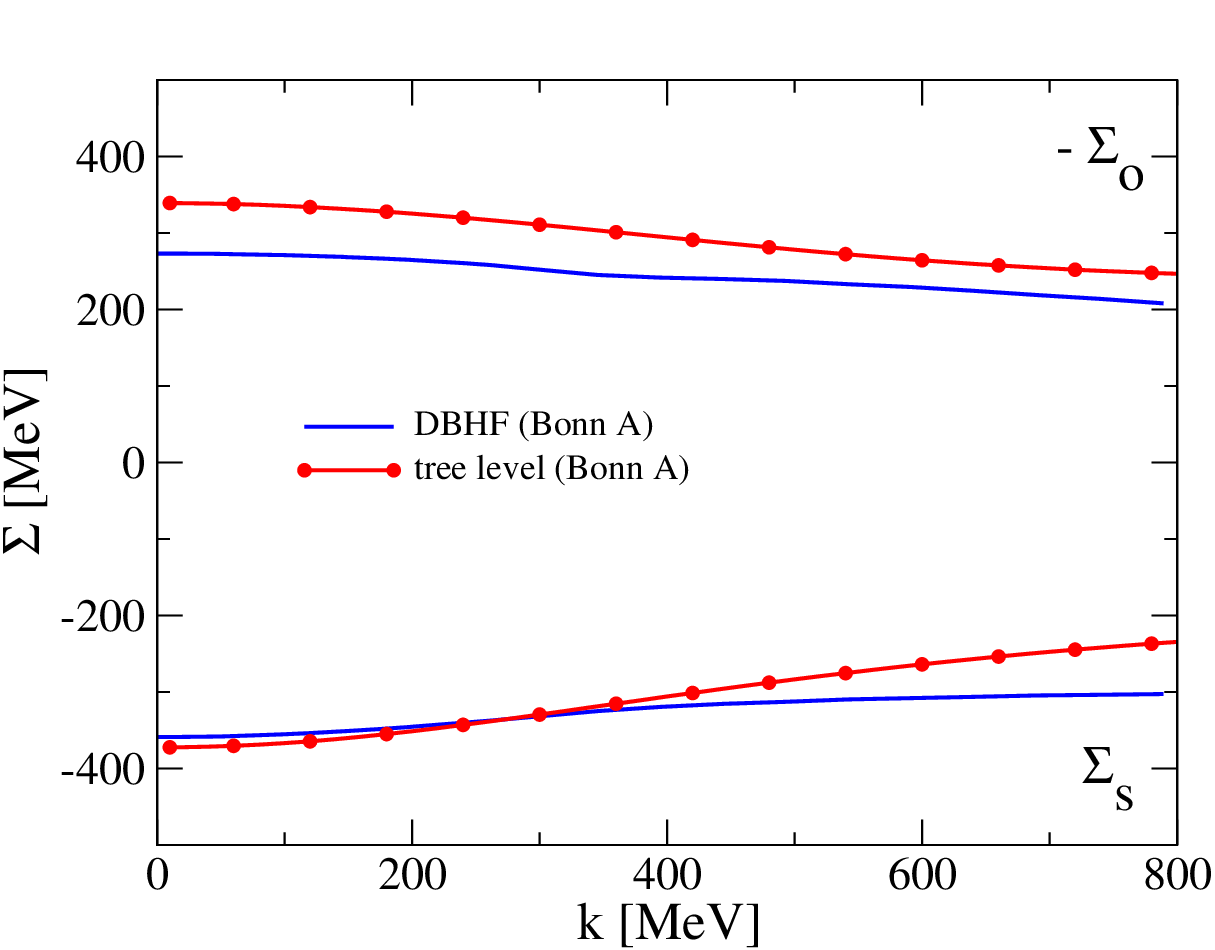}
\caption {(Color online) Tree level scalar and vector self-energy components 
in nuclear matter at $\kf =1.35~{\rm fm}^{-1}$ are compared to 
corresponding values from a full self-consistent relativistic 
Brueckner (DBHF) calculation. In both cases the Bonn A potential 
is used.
\label{sigma4} }
\end{center}
\end{figure}

In order to estimate the influence of short-range correlations and 
self-consistency, in Fig.~\ref{sigma4} the tree level result from 
Fig.~\ref{fig_sigma1} for Bonn A to a corresponding full DBHF calculation 
are compared 
at $\kf =1.35~{\rm fm}^{-1}$. For DBHF the approach of~\cite{gross99} is 
used (subtracted T-matrix in $pv$ representation). The DBHF calculation 
yields reasonable saturation properties with a binding energy of 
$E_{\rm bind} = -15.72$ MeV and a saturation density of 
$\rho =0.181~{\rm fm}^{-3}$~\cite{gross99}. It is no doubt that 
higher order correlations 
are essential for saturation of nuclear matter. The correlations lead 
to a general reduction of the vector self-energy by a shift of about 
70 MeV. Self-consistency and correlations also weakens the momentum 
dependence, in particular for $\Sigs$. However, except of the 
70 MeV shift of  $\Sigo$,  the absolute magnitude of the self-energies 
is not strongly modified in the realistic calculation. This means that one 
can expect that the large attractive scalar and repulsive vector mean 
fields will also persist for the other interactions when 
short-range correlations are accounted for in a full relativistic 
many-body calculation.  

\begin{figure}
\begin{center}
\includegraphics[width=0.75\textwidth] {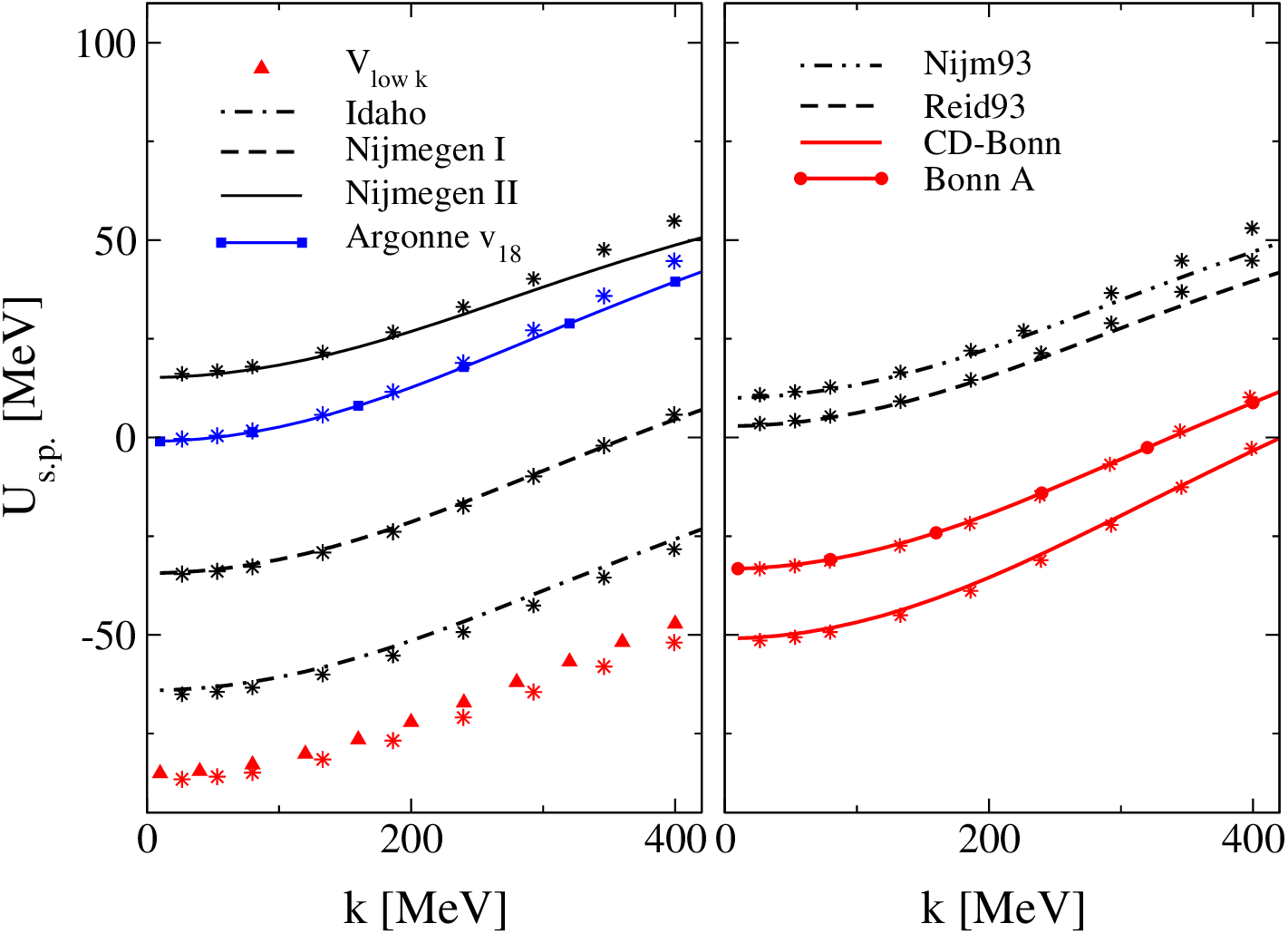}
\caption {(Color online) Single particle potential 
in nuclear matter at $\kf =1.35~{\rm fm}^{-1}$, determined from 
the tree level Born amplitudes of the various potentials. The 
single particle potential determined from the relativistic 
self-energy components after projection onto the covariant 
operator basis is compared to a non-relativistic calculation (stars) 
where partial wave amplitudes are summed up directly.  
\label{ufig1} }
\end{center}
\end{figure}

Fig.~\ref{ufig1} shows finally the single particle potential 
in nuclear matter at $\kf =1.35~{\rm fm}^{-1}$, determined from the 
relativistic self-energy components. The single particle potential is 
defined as the expectation value of the self-energy 

\begin{eqnarray}
   U_{\rm s.p.}(k,\kf ) = \frac{<u(k)|\gamma^0  \Sigma | u(k)>}
{< u(k)| u(k)>} =
\frac{M}{E({\bf k})} 
\, <{\bar u(k)}| \Sigma | u(k)>
\label{upot1}
\end{eqnarray}
and reads 
\begin{eqnarray}
U_{\rm s.p.} (k,\kf) = \frac{M}{E} \Sigs - \frac{ k_{\mu} \Sigma^\mu}{E} 
         = \frac{M \Sigs }{\sqrt{ {\bf k}^2 + M^{2}}} 
         - \Sigo + \frac{ \Sigv {\bf k}^2}
           {\sqrt{ {\bf k}^2  + M^{2}}}
\quad .
\label{upot2}
\end{eqnarray}
Eq.~(\ref{upot2}) represents the single particle potential at tree level, 
i.e., the expectation value of $\Sigma$ with the bare spinor basis. The 
next step towards a self-consistent treatment would be to use an in-medium 
spinor basis which includes the scalar and vector self-energy components via 
effective masses and effective four-momenta 
\beq
M^*(k,k_F)=M+\Sigs(k,k_F),\quad\quad k^{*}_\mu = k_\mu + \Sigma_\mu(k,k_F)~~. 
\label{effmass}
\eeq
This would, however, involve higher 
order corrections in the baryon density and is not intended in the 
present investigations which are restricted to leading oder. 

The single particle potential reflects the well known fact that 
 phase-shift equivalent two-body potentials which describe $NN$ 
scattering data with about the same accuracy~\cite{muether00}, can be 
rather different~\cite{muether00}. This can already be seen from 
Fig.~\ref{partialwaves} where the $^1S_0$ matrix elements of the various 
potentials are shown. The differences are mainly due to a different treatment 
of the short-range part of the nuclear interaction, 
i.e., the hard core which is not well constraint by 
scattering data. Thus the various potentials lead to 
about the same T-matrices when iterated in the Lippmann-Schwinger 
or Bethe-Salpeter equation. However, at tree-level the hard core 
contributes fully to $U_{\rm s.p.}$ which explains the shift of the 
various results in Fig.~\ref{ufig1}. 
Integrating out the high momentum components, e.g.\ by renormalization 
group methods, one arrives at equivalent low-momentum potentials $V_{\rm low~k}$ 
\cite{lowk03}. Since  $V_{\rm low~k}$ contains no significant contributions 
from the hard core it gives already at tree level
a realistic single-particle potential. The situation is similar for the 
chiral EFT N$^3$LO Idaho potential. As can be seen from 
Fig.~\ref{partialwaves}    
Idaho is rather close to  $V_{\rm low~k}$, not only in the  $^1S_0$ partial 
wave, and correspondingly  both lead to comparable potentials. However, 
the slight shift of about 10 MeV between $V_{\rm low~k}$ and Idaho 
reflects again the subtle cancellation effects between the large scalar/vector 
fields, since at the scale of the fields, Fig.~\ref{sigma3}, both lie 
practically on top of each other.

In the present context the  single particle potential serves as an 
important check of the whole procedure. In Fig.~\ref{ufig1} the single particle
potential $U_{\rm s.p.}$ is shown, calculated from Eq.~(\ref{upot2}), i.e.,
after projecting the $NN$ potentials 
from the partial wave basis onto the covariant operator basis, 
determining then the relativistic 
self-energy components and finally~$U_{\rm s.p.}$. Fig.~\ref{ufig1} 
includes also the results from a 'non-relativistic' calculation of 
$ U_{\rm s.p.}$  where the partial wave  amplitudes are 
directly summed up. To do so we used a non-relativistic 
Brueckner-Hartree-Fock program \cite{bhfmuether} and determined 
the single particle potential in Born approximation. The non-relativistic 
results are represented by stars in Fig.~\ref{ufig1} and shown up 
to a momentum of 400 MeV. This avoids distortions from non-relativistic 
kinematics which occur at higher momenta. 
At moderate momenta the non-relativistic and the relativistic 
calculations show an excellent agreement which demonstrates 
the accuracy of the applied projection techniques. One has thereby 
to keep in mind that $U_{\rm s.p.}$ originates in the relativistic approach 
from the cancellation 
of the two scalar and vector fields which are both of the order of 
about 400 MeV.  
\section{The structure of the self-energy from chiral EFT}
With the projection formalism at hand one is now able to investigate 
the connection between the appearance of the matter fields 
and chiral dynamics in more 
detail. It allows in particular a straightforward and transparent 
discussion of the contributions which arise at different orders in the 
chiral expansion of the $NN$ interaction, see Eqs.~(\ref{chlag}) and 
(\ref{eq_pot1}). Such an investigation allows also to build the bridge to 
the reduction of the in-medium quark condensates which is usually interpreted 
as a signature for a partial restoration of chiral symmetry. 
\subsection{Role of contact terms}
We are now in the situation 
to calculate the relativistic scalar and vector self-energies from 
a chiral EFT nucleon-nucleon potential order by order. For this purpose 
we apply again the chiral Idaho potential~\cite{machleidtpriv}. This 
allows to separate the contributions from different orders in the 
chiral expansion of the $NN$ interaction and provides a connection to 
the low energy constants (LECs) which appear at the different orders. 

\begin{figure}
\begin{center}
\includegraphics[width=0.75\textwidth] {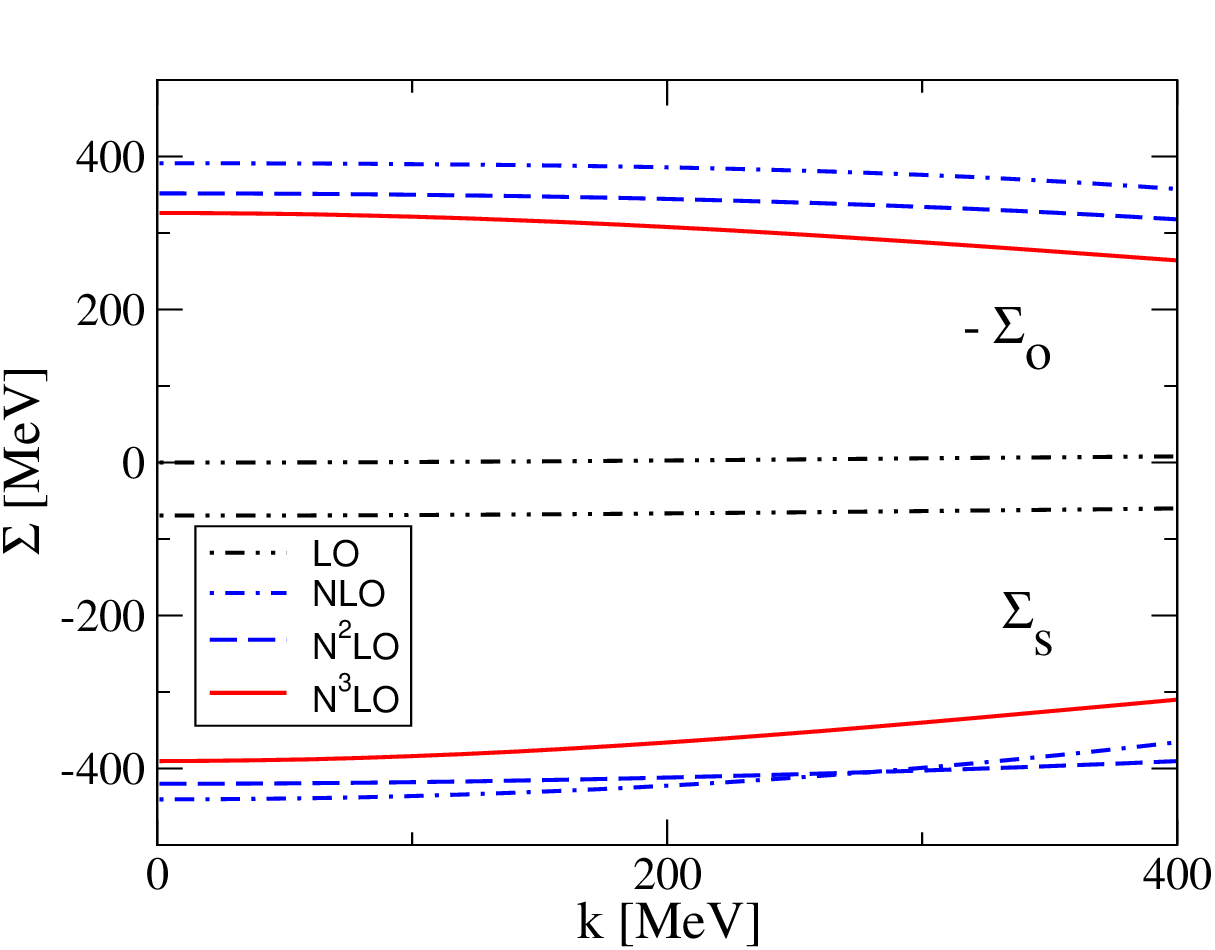}
\caption {(Color online) Tree level scalar and vector self-energy components 
in nuclear matter at $\kf =1.35~{\rm fm}^{-1}$ obtained with 
the chiral EFT  $NN$ interaction \protect\cite{entem03}. The 
fields obtained in leading order (LO) up to 
next-to-next-to-next-to-leading order (N$^3$LO) are shown.   
\label{sigorders} }
\end{center}
\end{figure}

Fig.~\ref{sigorders} shows 
the tree level results for the scalar and vector self-energy components 
in nuclear matter at $\kf =1.35~{\rm fm}^{-1}$ obtained in 
leading order (LO) up to next-to-next-to-next-to-leading order (N$^3$LO). 

To leading order the chiral $NN$ 
interaction does not generate significant mean fields. The scalar 
self-energy $\Sigs$ is of the order of about -70 MeV and the vector 
self-energy is practically zero. At LO only the static OPE and 
contact terms without derivatives appear which involve the operators 
$O_1$ and $O_2$ from the operator basis~(\ref{Pmom}). 
Hence at LO no pieces from 
vector exchange occur which would involve all operators $O_i, i=1..5$.  
The small scalar field means, 
on the other hand, that the nucleon mass $M^*$, Eq.~(\ref{effmass}), does not 
change significantly in matter to leading order in chiral EFT. The 
dominant contributions arise at next-to-leading order (NLO). NLO involves 
leading two-pion-exchange (2PE) and contact terms with two 
derivatives. The NLO contact terms contain the full operator 
structure $O_i$. At this level both, 
scalar and vector self-energy components of about $\mp 400$ MeV magnitude 
are generated. Also the signs, i.e., the attractive scalar and the 
repulsive vector mean field, are fixed at NLO. The higher orders, 
N$^2$LO and N$^3$LO provide corrections which tend to reduce the 
NLO result, are, however, moderate. N$^2$LO contains subleading 2PE
and no contact terms at all, while N$^3$LO contains sub-subleading 2PE, 
leading three-pion-exchange, corrections to OPE and 2PE and contact terms 
with four derivatives~\cite{epelbaum05}. 
\begin{table}
\begin{center}
\begin{tabular}{|l|c|c|c|c|c|c|}
\hline
   & ~~~$\Sigs$~~~ &~ $\Sigs^{(\pi)}$ &~ $\Sigs^{(\rm cont)}$ &~~$-\Sigo$~~~ & $-\Sigo^{(\pi)}$~ & $-\Sigo^{(\rm cont)}$ \\ 
\hline\hline
LO  & -64.76 & 17.14  & -81.9  & 4.49 & 19.02 & -14.53  \\
NLO  & -344.22 & 4.4  & -348.62  & 376.47 & 5.16 & 371.31  \\
N$^2$LO  & 2.06 & 2.06  & 0  & -41.92 & -41.92 & 0  \\
N$^3$LO  & 56.82 & -89.34  & 146.16  & -43.27 & 79.06 & -122.33  \\
\hline\hline
sum  & -350.1 & -65.74  & -284.36  & 295.77 & 61.32 & 234.45  \\
\hline
\end{tabular}
\end{center}
\caption{\label{tab1} Contributions from pion dynamics and 
 contact terms to the scalar and vector self-energy components (in MeV) 
which appear at different orders in the chiral expansion. The evaluation is 
performed at nuclear saturation density $\kf =1.35~{\rm fm}^{-1}$.}
\end{table}

In order to investigate the role of pion dynamics and that of  
contact terms in more 
detail, Table~\ref{tab1} contains the contributions which arise 
from pion dynamics $\Sigma^{(\pi)}$, i.e., OPE, 2PE, 3PE and corrections, and 
those from the contact terms $\Sigma^{(\rm cont)}$ separately. The 
contributions to the self-energy at a particular order is given 
by the sum  $\Sigma^{(\pi)} + \Sigma^{(\rm cont)}$, the full 
self-energy at a certain order $\nu$ is obtained by adding the contributions 
from the lower orders   
$\Sigma^{(\nu)}=\sum_{\lambda = 0}^\nu\Sigma^{(\lambda)}$. From Table~\ref{tab1} 
it becomes evident that the dominant contributions to the scalar and 
vector self-energy are generated by the contact terms which arise at 
next-to-leading order. At  N$^2$LO no contact terms occur in the chiral 
expansion. The  N$^3$LO contacts provide sizeable corrections to 
both, scalar and vector self-energy components and are of opposite sign 
than the NLO contributions. The contribution from pion dynamics to 
the self-energy components are found to be generally moderate. The 
largest contributions appear at N$^3$LO and are of opposite sign than 
those from corresponding contact terms. 

\begin{figure}
\begin{center}
\includegraphics[width=0.75\textwidth] {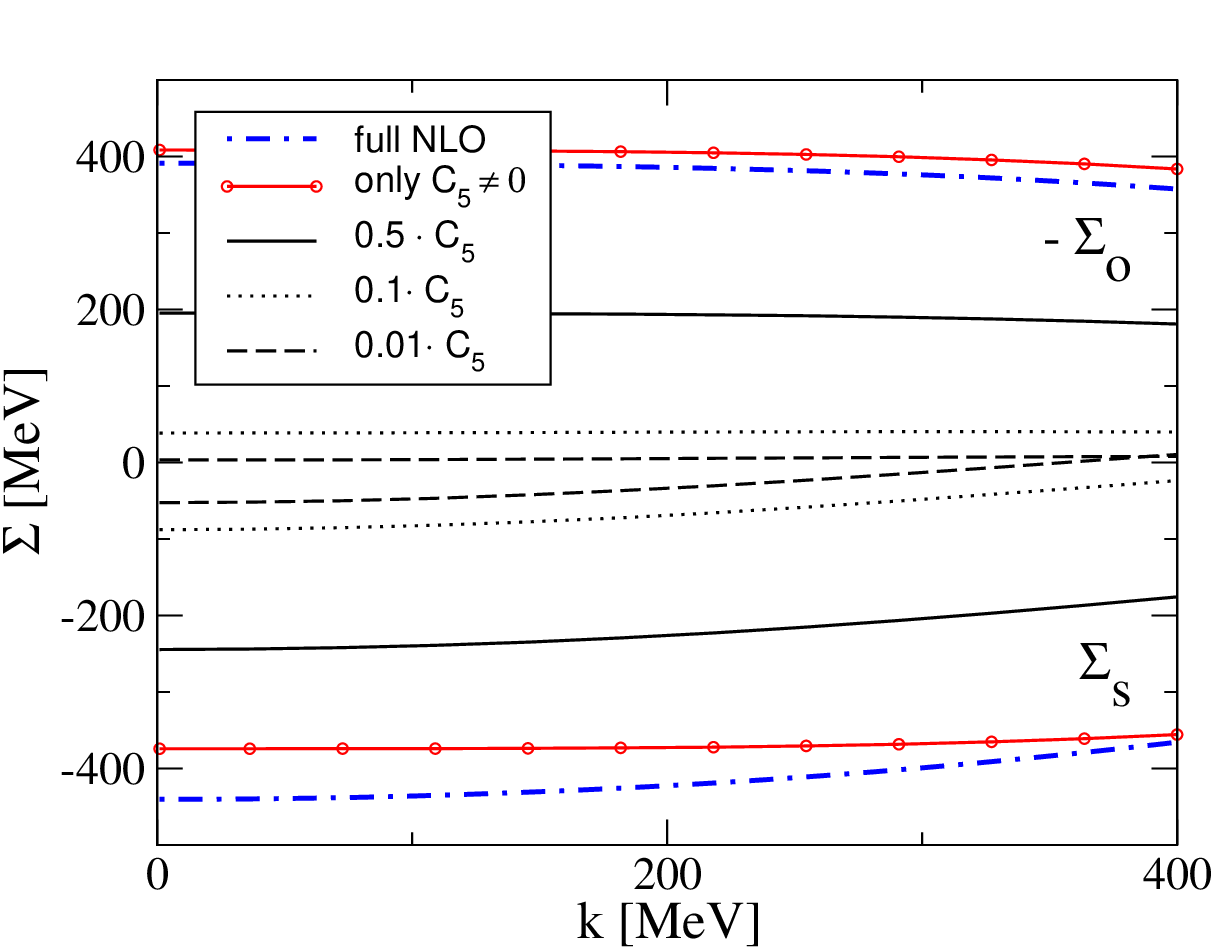}
\caption {(Color online) Influence of the $C_5$ low energy constant. 
The figure compares the self-energies at NLO to those where all contacts 
except of $C_5$ are switched off and those results where the strength 
of the $C_5$ parameter is varied. 
\label{c5dep} }
\end{center}
\end{figure}

Hence the reduction 
of the nucleon mass $M^* = M+\Sigs$ is driven by short-distance physics, 
dominantly by contact terms which occur at NLO. These are four-nucleon 
contacts with two derivatives. At this order the short-range spin-orbit  
interaction (proportional to $O_4$ in~(\ref{Pmom}))
\beq
iC_5
        (\mbox{\boldmath $\sigma$}_{1}+\mbox{\boldmath $\sigma$}_{2})
        \cdot({\bf q}\times{\bf q'} )   
\label{SO1}
\eeq
is generated. The appearance of large scalar/vector fields at NLO is 
therefore in perfect agreement with Dirac phenomenology where the 
large spin-orbit force is intimately connected to the appearance 
of the scalar/vector fields which are generated by short-range 
isoscalar scalar ($\sigma$) and vector meson ($\omega$) 
exchange~\cite{gross99,schiller01}. In EFT the 
strength of the short-range spin-orbit interaction is determined 
by the $C_5$ parameter which is given 
by a linear combination of the $^{3}P$-wave low energy constants 
(LECs)~\cite{entem03,epelbaum05}
\beq
C_5 = \frac{1}{16\pi}\left[ 2C_{3P0} + 3C_{3P1}-5C_{3P2}\right]~~.
\eeq
Hence the short-range spin-orbit interaction is dictated 
by $P$-wave $NN$ scattering. As shown by Kaiser~\cite{kaiser04} the 
large values of the $C_5$ parameter is in good agreement with corresponding 
values extracted from high precision OBE type potentials 
(Bonn, CD-Bonn, Nijm93, Nijmegen I,II) and from Argonne $v_{18}$ which 
are all in the range of $3C_5/8 \sim 80\div 90~{\rm MeV~fm}^5$. 
In~\cite{kaiser04} these values were 
also compared to purely phenomenological Skyrme type density 
functionals designed for nuclear structure 
calculations~\cite{reinhard04,skyrme04}. The values of the corresponding 
spin-orbit strength parameter $W_0$ in Skyrme models 
are also very close in magnitude, i.e. $3W_0/4 \sim 75\div 97~{\rm MeV~fm}^5$. 
 The contribution from chiral OPE to the spin-orbit terms in the 
density functional were found to be almost negligible (less than 1\%). 
The lowest order irreducible 
2PE which occurs at NLO in the chiral expansion provides  moderate 
corrections to the iso-scalar spin-orbit strength function 
whereas the iso-vector strength is more strongly affected 
(2PE contributions lead to a $\sim 30\%$ reduction)~\cite{kaiser04}. 
Thus the analysis of Kaiser is fully consistent with the small fields 
$\Sigs^{(\pi)}$ and $\Sigo^{(\pi)}$ of $\sim \mp 5$ MeV generated by 
pion dynamics at NLO, as observed within the framework of the present 
analysis. 

Fig.~\ref{c5dep} analyses the dependence of the fields on the value of 
the $C_5$ low energy constant in more detail. As already mentioned, 
at LO two contact terms ($C_1$ and $C_2$) appear and at NLO, respectively,  
5 contacts ($C_3$ to $C_7$). The figure contains the full NLO result, 
including contributions from LO and NLO pion dynamics and contacts 
and compares 
this to the case where all contacts which appear up to NLO were switched 
off except of the $C_5$ contribution. It contains in addition results 
with again all contributions, however, scaling the value of $C_5$ down 
to 50\%, 10\% and 0.1\%. It becomes evident that the large scalar and 
vector mean fields are a direct consequence of the large value of $C_5$.  
Chiral EFT is therefore not only 
in qualitative but quantitative agreement with the picture known from meson-exchange. 
In both cases the fields are related to short distance physics and their 
strength is dictated by $P$-wave $NN$ scattering data where 
the spin-orbit forces occur. 

\subsection{Connection to QCD sum rules}
In finite density QCD sum rules scalar and vector fields arise 
naturally from the structure of the quark propagator which is 
proportional to the corresponding condensates. As shown 
by Cohen et al.~\cite{cohen91} the quark correlation function 
can be expressed to leading order in terms of the scalar condensate 
$\langle\rho | {\bar q}q |\rho \rangle $ already present in vacuum,  
and the vector condensate $\langle\rho | q^\dagger q |\rho \rangle $ 
which is introduced by the breaking of Lorentz invariance due to the presence 
of the medium. The identification of the correlation function with the 
in-medium nucleon propagator of a dressed quasi-particle leads 
to scalar and vector self-energies 
$\Sigs$ and $\Sigo$ which are of the same order in the 
condensates~\cite{cohen91}
\beqa
\Sigs &=& - \frac{8\pi^2}{\Lambda^{2}_B} 
[\langle\rho | {\bar q}q |\rho \rangle  -\langle {\bar q}q \rangle]  
= - \frac{8\pi^2}{\Lambda^{2}_B} \frac{\sigma_N}{m_u+m_d} \rho_S 
\label{sum1}\\
-\Sigo &=& - \frac{64\pi^2}{3\Lambda^{2}_B} 
\langle\rho| {\bar q}\gamma_0 q |\rho \rangle = 
- \frac{32\pi^2}{\Lambda^{2}_B} \rho~~.
\label{sum1b}
\eeqa
These expression are of leading order in density. 
$\rho_S $ in (\ref{sum1}) is the scalar nucleon density, 
$f_\pi = 93$ MeV the weak pion decay constant 
and $m_{u,d}$ are the current quark masses of about $5\div 10$ MeV. 
The pion-nucleon sigma term 
$\sigma_N = \langle N| m_u {\bar u}u + m_d {\bar d}d|N\rangle $ 
is determined by the $u$ and $d$-quark content of the nucleon and 
represents the contribution from explicit chiral symmetry breaking to 
the nucleon mass through the small, but non-vanishing current quark masses. 
It has an empirical value of about $\sigma_N \simeq 50$ MeV.
The Borel mass scale $\Lambda_B \simeq 4\pi f_\pi  \simeq 1$ GeV 
is the generic low energy scale of QCD which  separates the 
non-perturbative from the perturbative regime. It coincides with the 
chiral symmetry breaking scale $\Lambda_\chi$ of ChPT. 
Applying Ioffe's formula~\cite{ioffe81} for the nucleon mass 
$M \simeq -\frac{8\pi^2}{\Lambda^{2}_B} \langle {\bar q}q \rangle$ one 
finally obtains the fields in the form~\cite{finelli} 
\beqa 
\Sigs (\rho) &=& - \frac{\sigma_N M}{m_\pi^2 f_\pi^2} \rho_S~~, 
\label{sum2}\\
-\Sigo (\rho) &=& \frac{4(m_u+m_d)M}{m_\pi^2 f_\pi^2} \rho~~.
\label{sum2b}
\eeqa 
However, the dependence of the 
nucleon mass in matter on the quark condensate is not as 
straightforward as expression (\ref{sum2}) suggests. 
Concerning the in-medium condensate one has carefully to distinguish 
between contributions from the pion cloud and those of non-pionic 
origin~\cite{birse96,chanfray01}. 

As pointed out by Birse~\cite{birse96} a naive direct dependence of the 
nucleon mass on the quark condensate through Eq.~(\ref{sum2}) 
leads to contradictions with chiral 
power counting. The contributions from low momentum virtual 
pions which enter the in-medium condensate should not contribute by the 
same amount to the change 
of the nucleon properties in matter. They can therefore 
not as easily be associated with a partial restoration of chiral symmetry 
as the mean field field approximation, Eqs.~(\ref{sum2}, \ref{sum2b}), would 
suggest. This problem has also been 
investigated by Chanfray et al.~\cite{chanfray01} in the framework 
of the linear sigma model. In their studies the authors were able to 
reconcile the phenomenology of Quantum Hadron Dynamics with chiral theory, 
in that case the linear sigma model. Their conclusion was that, in contrast to 
the scalar condensate $\langle\rho | {\bar q}q |\rho \rangle$ 
which is driven by the sigma field, i.e., the 
chiral partner of the pion, the lowering of the nucleon mass $M^*$ is driven 
by a chiral invariant scalar field which corresponds to fluctuation along the 
chiral circle. With other words, the condensate is to large extent reduced 
by the pion cloud surrounding the nucleons while the nucleon mass is not.

To set up the context for the following discussion, we shortly sketch 
the argumentation of Birse~\cite{birse96}:  
From Eq.~(\ref{sum2}) follows that the effective nucleon mass 
$M^* = M+ \Sigs (\rho)$ is directly proportional to the nucleon 
sigma term 
\beq
M^* = M\left( 1- \frac{\sigma_N }{m_\pi^2 f_\pi^2} \rho_S\right) ~~.
\label{mstar1}
\eeq 
The chiral expansion of the sigma term leads to \cite{gasser82}
\beq
\sigma_N = A m_\pi^2 - \frac{9}{16\pi}\left(\frac{g_{\pi NN}}{2M}\right)^2 m_\pi^3 + \dots
\label{sigma1}
\eeq
In the chiral limit the pion-nucleon coupling is connected to the axial 
vector coupling by the Goldberger-Treiman relation $g_{\pi NN}=g_A M/f_\pi$. 
The coefficient $A$ involves counter terms related to short-distance 
physics whereas the non-analytic ${\cal O}(m_\pi^3)$ term arises purely from long-distance 
physics of the pion cloud. Inserting~(\ref{sigma1}) into~(\ref{mstar1}) 
implies a dependence of the effective nucleon mass $M^*$ on the pion mass 
which is of order ${\cal O}(m_\pi)$. 

At the mean field level, i.e., in $T-\rho$ 
approximation, the scalar self-energy~(\ref{sigHF1}) is on the other hand 
given by the scalar forward scattering amplitude 
$ T_s({\bf q}=0)$ ( $ T_s({\bf q}=0)$ in~(\ref{trho}) 
corresponds to the direct amplitudes  $F_S$ and $g_S$ in~(\ref{ps-sc}) 
and~(\ref{pv_full}), respectively.)
\beq
\Sigs(\kf) = T_s({\bf q}=0)\rho ~~~. 
\label{trho}
\eeq
A comparison of Eq.~(\ref{trho}) with Eqs.~(\ref{mstar1}) and~(\ref{sigma1}) 
would imply that the scalar part of the forward scattering 
amplitude contains a constant and a term of order $m_\pi$. Such a dependence 
contradicts, however,  chiral power counting. In chiral EFT the leading 
term in the pion mass in the $NN$ interaction originates from the low 
energy expansion of the OPE and is of order 
${\cal O}(m_\pi^2)$~\cite{entem02,entem03,epelbaum05}. 
Hence the $NN$ interaction cannot 
contain a term directly proportional to $\sigma_N/f_\pi^2$.

\begin{figure}
\begin{center}
\includegraphics[width=0.75\textwidth] {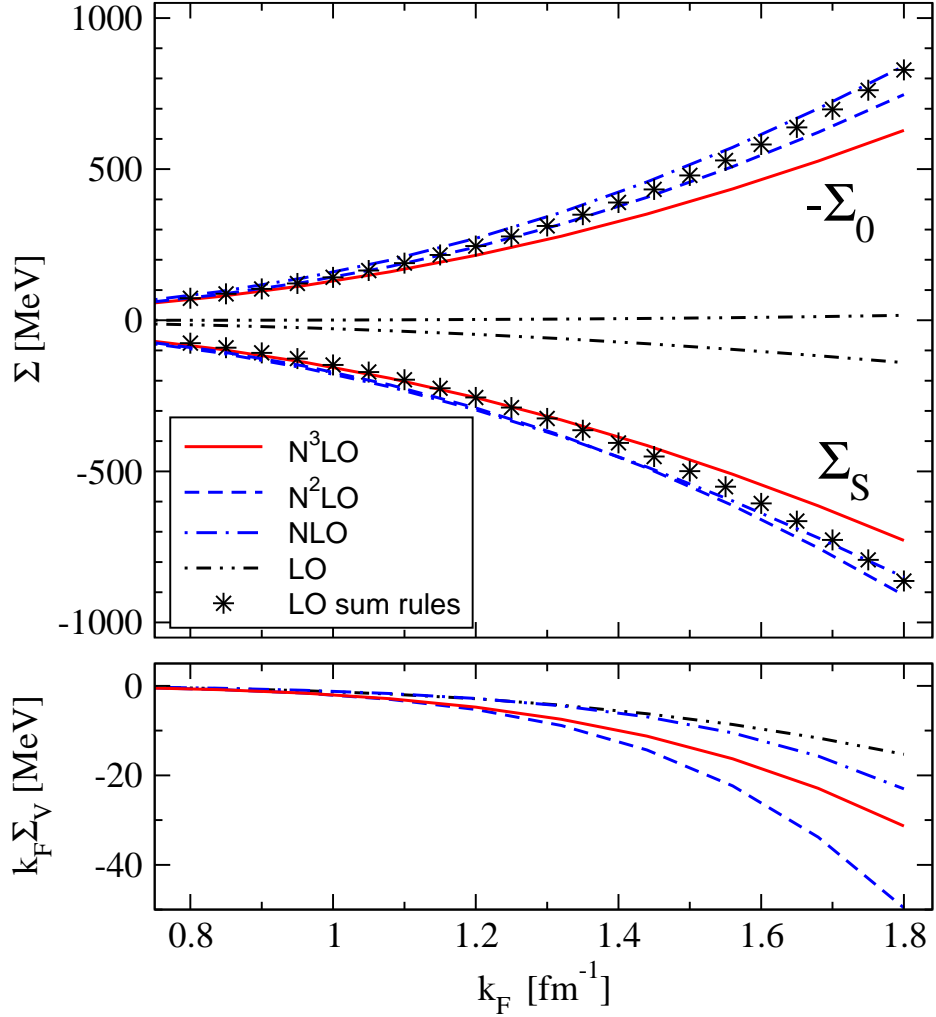}
\caption {(Color online) Density dependence of the tree level scalar and 
vector self-energy components in nuclear matter obtained with 
the chiral EFT $NN$ interaction \protect\cite{entem03}. The 
fields obtained in leading order (LO) up to 
next-to-next-to-next-to-leading order (N$^3$LO) are shown. The 
results from leading order QCD sum rules are shown as well.   
\label{sigorders2} }
\end{center}
\end{figure}

For the comparison of the sum rule predictions we turn to 
the density dependence of the self-energy. 
Fig.~\ref{sigorders2} shows the density dependence of 
the fields from the various orders. 
As in Fig.~\ref{sigma3}, the scalar $\Sigs$, 
time-like vector $\Sigo$ and spatial vector $\Sigv$ self-energies are 
determined at momentum $k=\kf$. The density dependence is shown up to 
$\kf =1.8~{\rm fm}^{-1}$ which corresponds to about 2.5 times nuclear 
saturation density. As can be seen from Fig.~\ref{sigorders2} the 
relative contributions from the various orders remain the same over the 
entire density range considered. For comparison the figure 
contains also the corresponding fields 
as predicted by leading order QCD sum rules, i.e., Eqs.~(\ref{sum2}) 
and~(\ref{sum2b}). 
For the evaluation of Eqs.~(\ref{sum2}) the empirical value of 
$\sigma_N = 50$ MeV has been chosen for the nucleon sigma term, 
$f_\pi = 93$ MeV and $(m_u+m_d)=12$ MeV. For the  evaluation of the 
scalar field in (\ref{sum2}) we have set 
the scalar density equal to the vector density, i.e., $\rho_s\simeq \rho$. 

Both, the QCD sum rule and the chiral EFT fields are well comparable 
in terms of a density expansion since both are obtained to leading 
order in density. In the case of the sum rules this corresponds to  
a Fermi gas of non-interacting nucleons. To go beyond the 
Fermi gas approximation would require to include higher order terms 
in the operator product expansion and the density expansion of the 
condensates~\cite{cohen91,drukarev91,kaempfer05}.  
In the EFT case higher orders  
in density can be introduced by a self-consistent dressing of the 
interaction (see discussion in Sec.~\ref{eossec}) and of course 
by  higher order terms in perturbation series which would finally 
end up in a full resummation of the Brueckner ladder diagrams.

At moderate nuclear densities the agreement between the QCD sum rules 
and N$^3$LO is quite remarkable. At higher densities the results from 
the sum rules tend to overshoot the  N$^3$LO values which is, 
however, not too astonishing since the relations (\ref{sum2}) are 
valid in the low density limit. 

In view of the fact that in chiral 
$NN$ dynamics the fields are dominantly generated by NLO contact terms, 
one could be tempted to interpret the present results in the way that 
the reduction of the quark condensates occurs at NLO in the chiral 
expansion. However, as discussed above 
such an interpretation is not straightforward. 
A closer inspection of the terms which drive the sum rule result 
reveals the following: the coefficient $A$ 
in (\ref{sigma1}) is related to the unknown coupling $C_1$ in the effective 
ChPT pion-nucleon Lagrangian~\cite{becher99}. Becher and Leutwyler 
extracted a value of $A=3.7~{\rm GeV}^{-1}$  fitting the elastic $\pi N$ 
scattering amplitude at threshold~\cite{becher01}. Inserting this value into 
the sum rule expression (\ref{mstar1}) corresponds to a scalar self-energy (at 
$\kf=1,35~{\rm fm}^{-1}$) of $\Sigs = -513$ MeV 
at order $m_{\pi}^0$. At order $m_{\pi}$, 
i.e., when the ${\cal O}(m_\pi^3)$ term in the expansion~(\ref{sigma1}) 
is included, the sigma term of 46.7 MeV is already close to its 
empirical value and a self-energy of $\Sigs = -340$ MeV is obtained. 
Although this value for $\Sigs$ is 
astonishingly close to the NLO result from chiral $NN$ scattering, 
one has to keep in mind that already the LO result 
is of order $m_{\pi}^2$ in the pion mass. 
In contrast to the sum-rule approach there appears no significant 
repulsive contribution from pion dynamics which would correspond to 
the ${\cal O}(m_\pi^3)$ term in~(\ref{sigma1}).

The present results are therefore 
in qualitative agreement with the findings of 
Refs.~\cite{birse96,chanfray01}, namely that long-distance physics related 
to pion dynamics plays only a minor role for the reduction of the nucleon 
mass in matter. Relating the in-medium nucleon mass 
to the in-medium scalar condensate through expression (\ref{mstar1}) 
one should be very careful. Although the sum rule mean fields, 
Eqs.~(\ref{sum1}) and~(\ref{sum1b}), provide a reasonable 
approximation to the mean fields from chiral EFT, 
both approaches do not reflect the same physical concepts. The sum rule 
approach assumes that the nucleon properties are determined by 
the interaction with the in-medium condensates while conventional 
many-body approaches assume that the in-medium properties are 
determined by the interaction between the nucleons.

\section{Equation of State}\label{eossec}
Until now all calculations in this paper have been performed at 
{\it tree level}. It is, however, a well known fact that a realistic 
description of nuclear dynamics requires correlations beyond Hartree-Fock. 
Short-range correlations are known to be essential for nuclear binding 
whenever realistic interactions are used. 
This leads in lowest order of the Brueckner hole-line expansion to the 
ladder approximation of the Bethe-Goldstone equation for the 
in-medium $G$-matrix~\cite{gammel}, or the Bethe-Salpeter 
equation in the relativistic case~\cite{anastasio83}. 
In contrast to non-relativistic BHF where the saturation points of  
isospin saturated matter are allocated on the so-called 
{\it Coester} line, the relativistic Dirac-Brueckner-Hartree-Fock 
approach leads to rather reasonable saturation 
properties~\cite{terhaar87a,brockmann90,gross99}. 
For a review see~\cite{fuchs06}. 

In Hartree-Fock the matter turns out 
to be unbound, in particular when high precision potentials with 
a relatively strong repulsive hard core are applied, e.g.\ OBE type 
potentials or Argonne $v_{18}$. The situation is qualitatively different 
for low momentum interactions ($V_{\rm low~k}$, Idaho N$^3$LO) where the hard 
core is strongly suppressed by the high momentum cut-offs. For these 
interactions isospin saturated nuclear matter collapses and 
Brueckner ladder correlations do 
not improve on this situation~\cite{kuckei02}. Here the matter has to be stabilised by 
the inclusion of repulsive three-body-forces~\cite{bogner05}. Doing so, there 
appears a strong cut-off dependence at tree-level which can be removed 
when the second order term of the Brueckner perturbation series is 
added. $V_{\rm low~k}$ in combination with three-body-forces does not 
require a full resummation of the ladder diagrams but can already 
be treated within second-order perturbation theory~\cite{bogner05}.

In the present work we do not aim for a fully realistic description of 
the nuclear many-body problem but restrict the investigations to the 
Hartree-Fock level. 
The tree-level results discussed up to now are of leading order in 
density $\rho$. Higher order corrections in density can be taken into 
account when the bare potential matrix elements are replaced by in-medium 
matrix elements $V\mapsto V^*$. In the relativistic approach such 
a treatment is well defined. It means to evaluate the corresponding Feynman amplitudes 
(\ref{vobe2}) through an in-medium spinor basis $u_{\lambda}^*(k)$ 
where the nucleons  are dressed by the self-energy. Such a treatment 
requires, however, a definite structure of the interaction 
which allows to evaluate corresponding in-medium amplitudes. 
It is therefore at present restricted to OBE-type potentials. 

The dressing of the interaction through the self-energy leads automatically 
to a self-consistency problem which is e.g.\ solved within DBHF. 
The higher order density dependences which are introduced by such a 
procedure are considered to be one of the essential reasons for the 
improved saturation behaviour of relativistic DBHF compared to 
non-relativistic BHF. In the following we will study the role 
of  self-consistency at the Hartree-Fock level. 

As already mentioned, 
in a relativistic framework one uses an in-medium spinor basis 
where the scalar and vector self-energy components from 
Sec.~\ref{selfenergyinmatter} enter via
effective masses and momenta, see Eq.~(\ref{effmass}).
Furthermore the spatial vector self-energy component is usually absorbed 
introducing reduced effective masses and momenta 
\beq
\tilde M^{\ast} = \frac{M^{\ast}}{ 1+\Sigv}, \quad\quad
\tilde k^{\ast}_{\mu} = \frac{k^{\ast}_{\mu}}{ 1+\Sigv}~~.
\eeq 
Hence the kinetic energy can be written as 
\beq
{\tilde k_0^{\ast}}={\tilde E^{\ast}}=\frac{E^{\ast}}{1+\Sigv}=
\sqrt{{\bf k}^2+{\tilde M^{\ast 2}}}
\eeq
and the in-medium spinors of helicity $\lambda$ are 
 given by
\beq                   \label{inmedium_ds}
u_{\lambda}^*(k) = \sqrt{\frac{\tilde E^{\ast} + \tilde M^{\ast}}
 { 2 \tilde M^{\ast}}                  }
\left( \begin{array}{c}  1  \\
\frac{2 \lambda |{\bf k}| }{\tilde E^{\ast} + \tilde M^{\ast}} 
\end{array} 
  \right) \chi_{\lambda}~~.
\eeq
Thus the effective mass $\tilde M^{\ast}$ 
introduces a density dependence into the interaction.  
The effective mass is, however,  in general not only density but also momentum 
dependent. Based on the observation that this 
explicit momentum dependence is moderate, it is 
usually neglected and $\tilde M^{\ast}$ is fixed at the reference point $|{\bf k}|=k_F$.
In the so-called {\it reference spectrum approximation} the reduced effective 
mass $\tilde M^{\ast}_F=\tilde M^{\ast}(|{\bf k}|=k_F,k_F)$
 serves as an iteration
parameter.
$\tilde M^{\ast}$ is then the solution of the non-linear equation 
\beq
\tilde M^{\ast} = M + \Sigs(k_F,\tilde M^{\ast}) 
- \tilde M^{\ast} \Sigv(k_F,\tilde M^{\ast})  \label{rsa}
\eeq
which follows from the formulae above. 
 Self-consistency is now achieved by determining for a given start value of 
$\tilde M^{\ast}$ the in-medium matrix elements 
$V^{JS}_{L',L}(\vecq^\prime,\vecq)$. 
Therefore the Lorentz invariant amplitudes $F_m^{\rm I}(|{\bf{q}}|,\theta)$ and
$g_m^{\rm I}(|{\bf{q}}|,\theta)$, Eqs.~(\ref{ps-sc}) and~(\ref{pv_full}),  
as well as the transformation matrix $C_{im}$ of Eq.~(\ref{matrix_eq}) 
depend on $\tilde M^{\ast}$ and the Fermi momentum $k_F$
since the plane-wave helicity states $|\lambda_1\lambda_2\,\vecq\rangle$ of 
Eq.~(\ref{expansion}) are now medium-dependent~(\ref{inmedium_ds}).
The next step is to compute the self-energy components $\Sigs$, $\Sigo$ 
and  ${\bf k}\Sigv$. Since the Dirac propagator~(\ref{propagator}) 
describes dressed quasi-particles now, also in~(\ref{sigHF1}), ~(\ref{sigHF2}) 
and~(\ref{sigHF3}) the 
mass $M$ and energy $E$ have to be replaced by the effective quantities 
$\tilde M^{\ast}~,\tilde E^{\ast}$. 
Finally the new $\tilde M^{\ast}$ is determined.
This iteration procedure is repeated until convergence is reached.

\begin{figure}
\begin{center}
\includegraphics[width=0.9\textwidth] {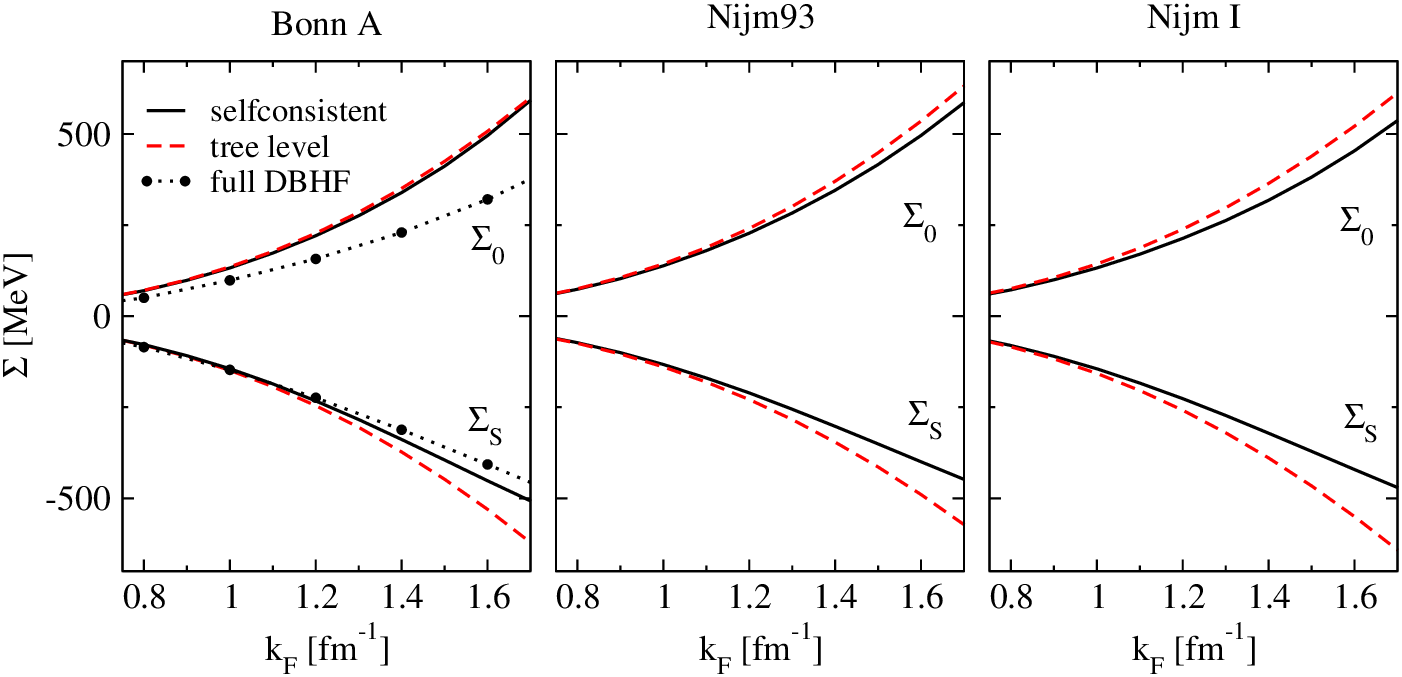}
\caption {(Color online) Comparison of the tree level scalar and vector 
self-energy components (dashed line) with self-consistent results (solid line).
Additionally a full self-consistent DBHF calculation is shown in the first 
graph denoted by dots. 
\label{self_cons} }
\end{center}
\end{figure}

In Fig.~\ref{self_cons} the results for the self-consistently calculated 
self-energy components $\Sigs$ and $\Sigo$ for  
Bonn A, Nijm 93 and Nijmegen I are shown as a function of the Fermi momentum 
and compared to the tree level results from Fig.~\ref{sigma3}. For the 
Bonn A case the result of a  full self-consistent DBHF calculation is 
shown as well~\cite{gross99}. From this figure two features can be observed: 
the higher order density dependences which are introduced by the 
dressing of the potential affect mainly the scalar part of 
the self-energy. The modifications of $\Sigo$ are moderate while 
$\Sigs$ is significantly reduced. The short range 
ladder correlations included in the full DBHF calculation (Bonn A) influence  
the self-energy in an opposite way. The deviations of $\Sigs$ from the 
self-consistent HF result are rather small, however, the vector component 
gets now strongly suppressed. This fact is understandable since 
the ladder correlations prevent the two-nucleon wave functions from 
too strong overlap with the hard core. In OBE 
potentials the hard core 
is mainly mediated by vector $\omega$-exchange and determines 
thus the vector self-energy 
component. 

With the self-consistent Hartree-Fock self-energies at hand one 
can now determine the equation of state (EOS). Like in DBHF the EOS, i.e., the 
energy per particle is defined 
as the kinetic plus half of the potential energy
\begin{eqnarray}                                      \label{eq26}
& &  {\rm E / A }   =   \frac{1}{\rho} \; \sum_{{\bf k},\lambda} 
< \overline{u}_{\lambda}^*({\bf k})|
\mbox{\boldmath$ \gamma $ \unboldmath} \cdot {\bf k} + M +
\frac{1}{2}\Sigma(k)|u_{\lambda}^*({\bf k}) > \frac{{\tilde M^{\ast}}}{{\tilde 
E^{\ast}}} - M    
                                                          \\ \label{eq27} 
& &  =  \frac{1}{\rho} \int_F \frac{d^3{\bf k}}{2\pi^3} 
    \left[ ( (1 + \Sigv(|{\bf k}|)){\tilde E^{\ast}} - \Sigma^0(|{\bf k}|) ) 
-\frac{1}{2{\tilde E^{\ast}}}( \Sigs(|{\bf k}|) {\tilde M^{\ast}} - \Sigma_{\mu}(|{\bf k}|) {\tilde k^{\ast\mu}} ) \right] - M 
\end{eqnarray} 
with the self-consistent spinors $u_{\lambda}^*$ from Eq. (\ref{inmedium_ds}).

\begin{figure}
\begin{center}
\includegraphics[width=0.9\textwidth] {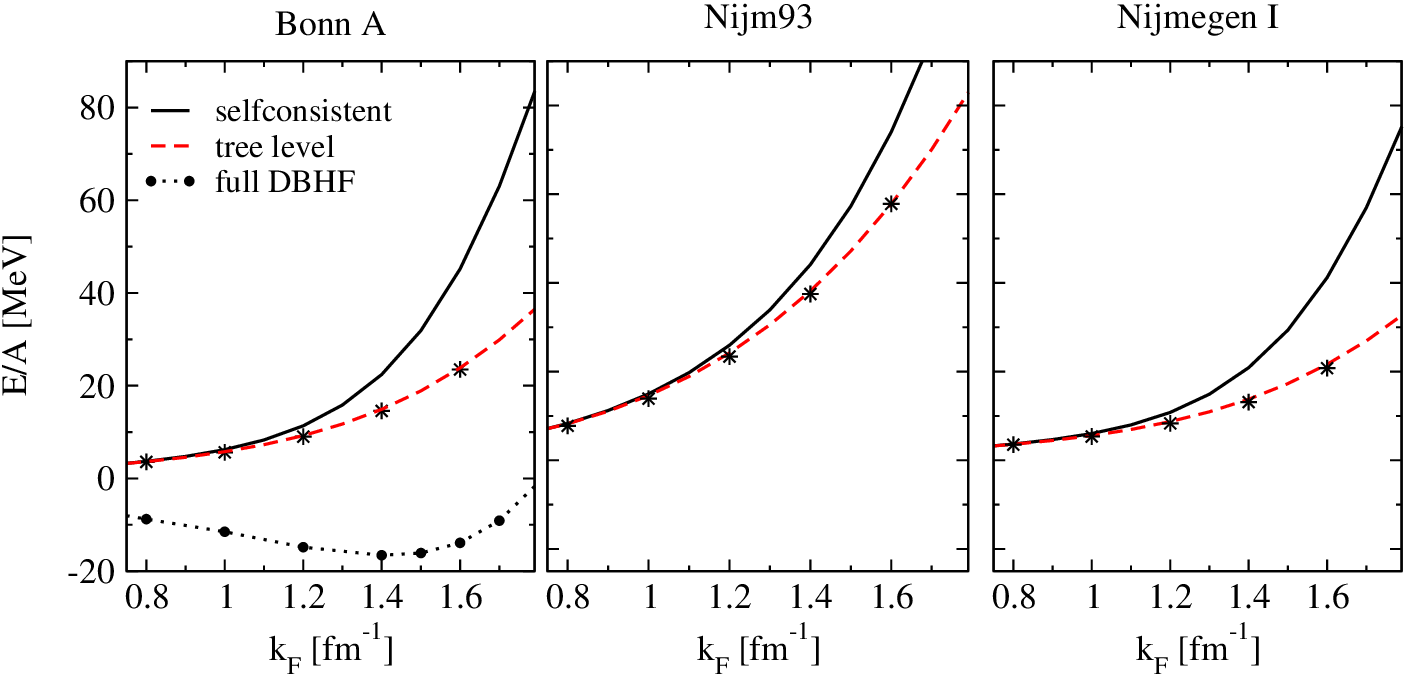}
\caption {(Color online) Hartree-Fock calculation of the nuclear equation 
of state, i.e., energy per particle E/A as a function of the Fermi momentum 
$k_F$ for three different potentials. The dashed line indicates a tree level 
calculation and the solid line represents a self-consistent Hartree-Fock 
calculation, i.e., higher order corrections in density are included.   
\label{EOS} }
\end{center}
\end{figure}

In Fig.~\ref{EOS}, we present the self-consistent Hartree-Fock results 
for the energy per 
particle in symmetric nuclear matter calculated from the Bonn A, Nijm93, 
Nijmegen I potentials as a function of the Fermi momentum $k_F$ which is a
measure for the density $\rho=2/(3\pi^2)k_F^3$.

Also a non-self-consistent calculation is shown (dashed line) where 
the energy per particle is given by
\beq
 {\rm E / A }=\frac{1}{\rho} \int_F \frac{d^3{\bf k}}{2\pi^3} 
    \left[ \frac{k^2}{2M}+\frac{1}{2}\,U_{\rm s.p.} (k,\kf) \right] 
\eeq
with $U_{\rm s.p.} (k,\kf)$  as defined in Eq.~(\ref{upot2}). In this case one
 obtains the same result as in a non-relativistic Hartree-Fock 
calculation (denoted by stars in Fig.~\ref{EOS}). 
The latter demonstrates again the numerical accuracy of the procedures.

For the Bonn A case again the EOS from the full DBHF calculations is 
shown as a reference~\cite{gross99}. It is clear that ladder correlations and 
other in-medium effects such as Pauli-blocking of intermediate states 
in the Bethe-Salpeter equation are responsible for nuclear saturation. 
The relatively moderate deviations from self-consistent Hartree-Fock at 
the scale of the self-energies in Fig.~\ref{self_cons} are essential at the 
scale of the binding energy. Like in relativistic mean field theory of 
QHD subtle cancellation effects in the large scalar and vector fields 
are responsible for nuclear binding. 

The higher order density dependences introduced via the dressing of the 
bare interaction $V$ lead to significantly more repulsion at the level of 
the EOS. This is a direct consequence of the reduced attractive scalar 
field (see Fig.~\ref{self_cons}). Thus Fig.~\ref{EOS} serves also as a 
demonstration for the success of DBHF compared to BHF what concerns 
the quantitative description of nuclear saturation: In particular for 
modern high precision potentials such as Bonn, Nijmegen or Argonne 
$v_{18}$ the BHF approach leads to strong over-binding and too high 
saturation densities. The additional repulsion introduced by 
higher order terms in density through the dressed potentials 
shifts the corresponding saturation points towards the 
empirical region~\cite{brockmann90,honnef,fuchs06}. We want to stress 
that the density dependence of the dressed potential $V^*$ 
should not be mixed up with 
the density dependence of the $G$-matrix. The latter originates from the 
dressed two-nucleon propagator and the Pauli-operator in the Bethe-Goldstone 
(or Bethe-Salpeter) equations while $V^*$ enters into the  Bethe-Salpeter 
for iteration. In 
non-relativistic BHF or variational calculations~\cite{zuo04,akmal98} 
a non-linear density dependence which improves
 the saturation behaviour is usually introduced through net repulsive 
three-body-forces. In such a treatment the dependence on the third particle 
is integrated out such that one is left with an additional effective 
density dependent two-body force which acts in a similar way as a dressing 
of the two-body interaction. In this context one should mention 
that a dressing of the interaction has also more subtle consequences 
when iterated in the Bethe-Salpeter equation. It leads e.g.\ to a 
quenching of the second order OPE exchange \cite{banerjee02} which plays an 
essential role for saturation in non-relativistic approaches.

In summary, one could expect that a dressing of the interaction would 
allow to comply with weaker three-body forces which may in particular be 
of interest concerning the application of low momentum EFT potentials to 
the nuclear many-body problem. 
As the studies of Bogner et al.~\cite{bogner05} have demonstrated, 
$V_{\rm low~k}$ 
requires rather strong three-body forces in order to stabilise nuclear 
matter. There the strength of the three-body contributions has already 
been pushed to its limits. Although a dressing of the interaction will 
probably not be possible for  $V_{\rm low~k}$ due to the partially non-analytic 
structure of the potential, it may be a promising perspective for the 
application of other EFT potentials, e.g.\ the chiral N$^3$LO. 

\section{Summary}
The appearance of large scalar and vector fields is a well established 
feature of relativistic nuclear dynamics. The  saturation mechanism of 
nuclear matter or the single particle potential in finite nuclei are 
obtained by subtle cancellation effects between large attractive scalar 
and repulsive vector fields. These fields occur already at tree level and 
do not change too much when realistic many-body calculations are performed.    
Full self-consistent Brueckner calculations which account  
for short-range ladder correlations lead to mean fields of similar size, i.e., 
of several hundred MeV magnitude.   
The size of the scalar-vector fields coincides with the values 
derived from relativistic mean field phenomenology by fits to 
finite nuclei. Alternatively, QCD sum rules come to the same results.

The present work addresses the question about the origin of 
these fields. When the nucleon-nucleon interaction is described 
within the framework of a meson exchange picture, the situation 
is rather clear. The Lorentz character of the mesons determines 
automatically the Lorentz character of the interaction at the 
corresponding scale: the short-range repulsion is due to vector 
exchange ($\omega, \rho$) while the intermediate range attraction 
originates from scalar exchange ($\sigma$).  As a direct consequence 
this leads to the existence of large scalar and vector mean fields 
in nuclear matter. However, these fields are not observables. 
It is therefore a fundamental questions of 
nuclear physics whether the appearance of large scalar/vector 
fields is intimately connected to the meson exchange 
picture or if it is a general consequence of the 
vacuum $NN$ interaction.  

To address the question in a model independent way, 
we based the present study on a broad set of modern high precision  
$NN$ potentials: Bonn, CD-Bonn, Nijmegen, Argonne $v_{18}$, 
Reid93, Idaho N$^3$LO and $V_{\rm low~k}$. Except the fact that 
all these potentials fit $NN$ scattering data with high accuracy they
are based partially on quite different theoretical concepts, i.e., 
the traditional meson exchange picture (Bonn, CD-Bonn, Nijmegen), 
a purely phenomenological philosophy (Argonne $v_{18}$, Reid93) 
or QCD inspired effective field theory approaches 
(Idaho N$^3$LO, $V_{\rm low~k}$).

For this purpose the potentials were projected on a relativistic 
operator basis in Dirac space. This was achieved using standard projection 
techniques which transform from a partial wave basis, i.e., the basis 
where the potentials are originally given, to the basis 
of covariant amplitudes in Dirac space. The idea behind this approach is that 
both, relativistic and non-relativistic descriptions of the $NN$ interaction 
have common features, i.e., they are based on a certain operator structure 
in spin-isospin space and invoke certain scales: the long-range part  
of scale $m_\pi$, essentially given by one-pion exchange, 
the intermediate range attraction and the  
short-range repulsion. In the meson exchange picture the various 
scales are associated with the meson masses which mediate the 
interaction. The various approaches can now 
be compared at the level of these  covariant amplitudes where 
we observe a remarkable agreement 
between the meson exchange potentials (Bonn, CD-Bonn, Nijmegen), 
the phenomenological non-relativistic potentials  
(Argonne $v_{18}$, Reid93) and the EFT potentials 
(Idaho N$^3$LO, $V_{\rm low~k}$).

Moreover, this procedure allows now to calculate the relativistic 
self-energy operator in nuclear matter. 
The key result of the present investigations is the tree level self-energy 
in nuclear matter. The structure of the nucleon-nucleon interaction enforces 
the existence of large scalar and vector fields. This is found to be a model 
independent fact, true for all types of interactions which have been
considered. The scale of these fields is set at tree level. Although 
essential for nuclear binding and saturation, higher order correlations, 
in particular short-range correlations, change the size of the fields by 
less than 25\%. The magnitude of the tree-level fields 
is very similar to that predicted by relativistic 
mean field phenomenology and  relativistic many body calculations. 

The connection to QCD as the underlying theory of strong interactions 
is established by chiral effective theory. EFT nucleon-nucleon
 potentials are derived 
from a systematic expansion of an effective Lagrangian which respects 
the basic symmetries of QCD. Chiral EFT is considered as the exact mapping 
of QCD on effective hadronic degrees of freedom in the non-perturbative 
regime. Subjecting the chiral  N$^3$LO Idaho potential to the present 
projection scheme 
we can make the following statements: In nuclear matter scalar and vector 
mean fields of the same sign and magnitude are generated as by the 
meson exchange or phenomenological potentials. These fields are generated  
by contact terms which occur at next-to-leading order in the chiral 
expansion. These are four-nucleon contact terms with two derivatives 
which generate the short-range spin-orbit interaction. The strength 
of the corresponding low energy constants, in particular those connected 
to the spin-orbit force, is dictated by $P$-wave $NN$ scattering data.  
Pion dynamics as well as LO and N$^3$LO contacts provide 
only corrections to the fields generated by the NLO contact terms. 
EFT is therefore in perfect agreement with Dirac phenomenology where 
it is known since a long time that the large scalar/vector fields are 
generated by the short-range vector ($\omega$) and 
scalar ($\sigma$) mesons which are connected intimately to the large 
spin-orbit interaction. We conclude that this is a direct consequence 
of  $P$-wave $NN$ scattering.

Like in OBE models and RMF theory, in EFT the reduction 
of the nucleon mass $M^* = M+\Sigs$ is driven by short-distance physics.  
Long-distance physics from virtual pions, i.e. the non-analytic term in 
the expansion of $\sigma_N$ gives a sizable contribution to the modification 
of the in-medium quark condensate. Such contributions are, however, 
found to play only a minor 
role for the reduction of the nucleon mass. Nevertheless, at moderate 
nuclear densities the N$^3$LO scalar and vector fields 
agree almost perfectly with the prediction from leading order QCD 
sum rules. For future perspectives chiral EFT in combination 
with projection techniques may allow 
to determine the relativistic anti-proton potential in matter 
in a model independent way. Here the meson-exchange picture predicts 
a change in sign of the vector field due to g-parity and hence an 
extremely deep attractive potential. 
Such investigations in particular will 
be interesting in view of the forthcoming anti-proton facilities, 
e.g. Panda at FAIR~\cite{peters06}. 

Finally we investigated implications of higher order corrections 
in density on the nuclear EOS. A dressing of the potential through 
self-consistently determined self-energies leads already at the 
Hartree-Fock level to significantly more repulsion in the EOS. 
At present these 
investigations  were restricted to OBE type potentials. But to 
include such higher order terms in density might open a promising 
perspective also for EFT potentials when applied to the nuclear 
many-body problem.  

\acknowledgments
We like to thank D. Entem for providing the Idaho N$^3$LO program 
package, H. M\"uther for providing his Brueckner-Hartree-Fock 
code and R. Machleidt for providing a version of the Idaho N$^3$LO program 
which allows to easily separate the contributions from the various orders. 
This work was supported by the DFG under contract GRK683 (European Graduate 
School T\"ubingen-Basel-Graz).



\newpage
\begin{thebibliography}{99}

\bibitem{sw86} 
        B. D. Serot, J. D. Walecka,  
        Advances in  Nuclear Physics, {\bf 16}, 1, 
        eds. J. W. Negele, E. Vogt, (Plenum, N.Y., 1986). 
 
\bibitem{rmf} 
P. Ring, Prog. Part. Nucl. Phys. {\bf 73}, 193  (1996);  
Lect. Notes Phys. {\bf 641}, 175 (2004). 

\bibitem{arima69}  
A. Arima, M. Harvey, K. Shimizu, Phys. Lett. {\bf B30}, 517 (1969).

\bibitem{ginocchio97}
J.N. Ginocchio, Phys. Rev. Lett. {\bf 78}, 436 (1997).

\bibitem{ring96b} 
A.V. Afanasjev, J. K\"onig, P. Ring, Phys. Lett. {\bf B367}, 11 (1996).  

\bibitem{cohen91}
 T.D. Cohen, R.J. Furnstahl, D.K. Griegel, 
Phys. Rev. Lett. {\bf 67}, 961 (1991); Phys. Rev. C {\bf 45}, 1881  (1992).

\bibitem{drukarev91}
E.G. Drukarev, E.M. Levin, Prog. Part. Nucl. Phys. {\bf 27}, 77  (1991).

\bibitem{ioffe81}
B.L. Ioffe, Nucl. Phys. {\bf B188}, 317  (1981).

\bibitem{finelli} 
P. Finelli, N. Kaiser, D. Vretenar, W. Weise,  
Eur. Phys. J. A {\bf 17}, 573   (2003); Nucl. Phys. {\bf A735}, 449 (2004). 

\bibitem{gammel}
K.A. Brueckner, J.L. Gammel,  Phys. Rev. {\bf 107}, 1023 (1958).

\bibitem{anastasio83}
M. R. Anastasio, L. S. Celenza, W. S. Pong, and C. M. Shakin,
Phys. Rep. {\bf 100}, 327 (1983).

\bibitem{terhaar87a}
B. ter Haar and R. Malfliet,
Phys. Rep. {\bf 149}, 207 (1987).

\bibitem{brockmann90}
R. Brockmann and R. Machleidt, Phys. Rev. C {\bf 42}, 1965 (1990).

\bibitem{gross99}
T. Gross-Boelting, C. Fuchs, and Amand Faessler,
Nucl. Phys. {\bf A648}, 105 (1999).

\bibitem{dalen04}
E. van Dalen, C. Fuchs, A. Faessler, Nucl. Phys. A 744 (2004) 227;
Phys. Rev. Lett. {\bf 95}, 022302 (2005); 
Phys. Rev. C \textbf{72}, 065803 (2005). 


\bibitem{honnef}
C. Fuchs, Lect. Notes Phys. {\bf 641}, 119  (2004).

\bibitem{bonn}
R. Machleidt, K. Holinde, Ch. Elster, Phys. Rep.  {\bf 149}, 1 (1987)

\bibitem{av18}
R.B. Wiringa, V.G.J. Stoks, R. Schiavilla, 
Phys. Rev. C {\bf 51}, 38 (1995).

\bibitem{nijmI_II}
V.G.J. Stoks, R.A.M. Klomp, C.P.F. Terheggen, and J.J. de Swart 
Phys. Rev. C {\bf 49}, 2950 (1994).

\bibitem{nijmegen}
V.G.J. Stoks, R.A.M. Klomp, M.C.M. Rentmeester, J.J. de Swart 
Phys. Rev. C {\bf 48}, 792 (1993).

\bibitem{nijm78}
M.M. Nagels, T.A. Rijken, J.J. de Swart 
Phys. Rev. D {\bf 17}, 768 (1978).

\bibitem{entem02} 
D.R. Entem, R. Machleidt, Phys. Rev. C {\bf 66}, 014002 (2002).

\bibitem{entem03} 
D.R. Entem, R. Machleidt, Phys. Rev. C {\bf 68}, 041001 (2003).

\bibitem{epelbaum05}
E. Epelbaum, W. Gl\"ockle, U.-G. Meissner, 
Nucl. Phys. {\bf A747}, 362 (2005). 

\bibitem{lowk03}
S.K. Bogner, T.T.S. Kuo, A. Schwenk, 
Phys. Rept. {\bf 386}, 1 (2003). 

\bibitem{tjon85a}
J. A. Tjon and S. J. Wallace,
Phys. Rev. C {\bf 32}, 267 (1985).

\bibitem{horowitz87}
C. J. Horowitz and B. D. Serot,
Nucl. Phys. {\bf A464}, 613 (1987).


        \bibitem{cdbonn}
R. Machleidt, Phys. Rev. C {\bf 63}, 024001  (2001).

\bibitem{machleidt01}
R. Machleidt, I. Slaus, J. Phys. {\bf G27}, R69 (2001).

\bibitem{muether00} 
H. M\"uther, A. Polls, Prog. Part. Nucl. Phys. {\bf 45}, 243 (2000). 


\bibitem{machleidt93}
        R. Machleidt, in: Computational Nuclear Physics 2 --- Nuclear
        Reactions, K. Langanke, J. A. Maruhn, and S. E. Koonin, eds.
        (Springer, New York, 1993), Chapter 1, p. 1.

\bibitem{Wei90} 
S. Weinberg, Phys.\ Lett.\ B {\bf 251}, 288 (1990);
Nucl.\ Phys.\ {\bf B363}, 3 (1991).

\bibitem{Kai01} 
N. Kaiser, Phys.\ Rev.\ C {\bf 63}, 044010 (2001); 
Phys.\ Rev.\ C {\bf 64}, 057001 (2001); 
Phys. Rev. C {\bf 65}, 017001 (2002).

\bibitem{tjon85b}
J.A. Tjon and S.J. Wallace,
Phys. Rev. C {\bf 32}, 1667 (1985).

\bibitem{fuchs98}
C. Fuchs, T. Waindzoch, A. Faessler and D.S. Kosov,
Phys. Rev. C {\bf 58}, 2022 (1998).

\bibitem{erkelenz74}
K. Erkelenz, Phys. Rep. {\bf 13}, 191 (1974).

\bibitem{machleidt89}
R. Machleidt, Adv. Nucl. Phys. {\bf 19}, 189 (1989).

\bibitem{rose57}
M. Rose, Elementary Theory of Angular Momentum (Wiley, New York, 1957).

\bibitem{sehn97}
L. Sehn, C. Fuchs, and A. Faessler, Phys. Rev. C {\bf 56}, 216  (1997).


\bibitem{bhfmuether}
This code was provided by H. M\"uhter and used in the 
Born option. For each potential partial waves up to $J=9$ 
are taken into account.  

\bibitem{birse96}
M. Birse, Phys. Rev. C {\bf 53}, R2048 (1996).

\bibitem{chanfray01}
G. Chanfray, M. Ericson, P.A.M. Guichon, 
Phys. Rev. C63 (2001) 055202

\bibitem{gasser82}
J. Gasser, H. Leutwyler, Phys. Rep. {\bf 87}, 77 (1982).

\bibitem{machleidtpriv}
This investigation is based on a version of the chiral 
N$^3$LO Idaho potential which allows to seperate the contributions 
from different orders. The code was provided by R. Machleidt (privat 
communication). 

\bibitem{kaempfer05}
R. Thomas, S. Zschocke, B. K\"ampfer
Phys. Rev. Lett. {\bf 95}, 232301 (2005).
     
\bibitem{plohl06}
O. Plohl, C. Fuchs, E. van Dalen, Phys. Rev. C {\bf 73}, 014003 (2006)

\bibitem{schiller01}
E. Schiller and H. M\"uther, Eur. Phys. J.  A \textbf{11}, 15 (2001).

\bibitem{becher99}
T. Becher and H. Leutwyler,  Eur. Phys. J.  C \textbf{9}, 643 (1999).

\bibitem{becher01}
T. Becher and H. Leutwyler, JHEP {\bf 0106}, 017 (2001).

\bibitem{kuckei02}
J. Kuckei, F. Montani, H. M\"uther, A. Sedrakian, 
Nucl. Phys. {\bf A723}, 32 (2003).

\bibitem{bogner05}
S.K. Bogner, A. Schwenk, R.J. Furnstahl, A. Nogga, 
Nucl. Phys. {\bf A763}, 59 (2005).
     
\bibitem{kaiser04}
N. Kaiser, Phys. Rev. C {\bf 70}, 034307 (2004)


\bibitem{reinhard04} 
P.-G. Reinhard, M. Bender,  
Lect. Notes Phys. {\bf 641}, 249 (2004). 

\bibitem{skyrme04}
B. Cochet, K. Bennaceur, J. Meyer, P. Bonche, T. Duguet, 
Int. J. Mod. Phys. {\bf E13}, 187 (2004).

\bibitem{fuchs06}
C. Fuchs, H.H. Wolter, nucl-th/0511070, to appear in 
Eur. Phys. J. A. 

\bibitem{zuo04} 
X.R. Zhou, G.F. Burgio, U. Lombardo, H.-J. Schulze, W. Zuo,  
Phys. Rev. C {\bf 69}, 018801 (2004). 

\bibitem{akmal98} 
A. Akmal, V.R. Pandharipande, D.G. Ravenhall,  
Phys. Rev. C {\bf 58}, 1804 (1998). 

\bibitem{banerjee02} 
M.K. Banerjee, J.A. Tjon, Phys. Rev. C {\bf 58}, 2120 (1998),  
Nucl. Phys. A {\bf 708}, 303 (2002). 
 
\bibitem{peters06}
K. Peters, Nucl. Phys. B (Proc. Suppl.) {\bf 154}, 35-41 (2006)
\end{thebibliography}
\end{document}